\documentclass[twocolumn]{aastex61}
\pdfoutput=1 %for arXiv submission
\usepackage{amsmath,amstext}
\usepackage[T1]{fontenc}
\usepackage{apjfonts} 
\usepackage[figure,figure*]{hypcap}
\usepackage{longtable}
\usepackage{comment}

\newcommand{\HII}{H\,\textsc{ii}}
\newcommand{\HI}{H\,\textsc{i}}
\newcommand{\HeI}{He\,\textsc{i}}
\newcommand{\NII}{N\,\textsc{ii}}
\newcommand{\OII}{O\,\textsc{ii}}
\newcommand{\OIII}{O\,\textsc{iii}}
\newcommand{\SII}{S\,\textsc{ii}}
\newcommand{\NeIII}{Ne\,\textsc{iii}}

\shorttitle{Searching for the Lowest Metallicity Galaxies}
\shortauthors{Hsyu et al.}

\begin{document}

\title{Searching for the lowest metallicity galaxies in the local universe}

\author{Tiffany Hsyu}
\affiliation{Department of Astronomy \& Astrophysics, University of California Santa Cruz, 1156 High Street, Santa Cruz, CA 95060}
\author{Ryan J. Cooke}
\affiliation{Centre for Extragalactic Astronomy, Department of Physics, Durham University, South Road, Durham DH1 3LE, UK}
\author{J. Xavier Prochaska}
\affiliation{Department of Astronomy \& Astrophysics, University of California Santa Cruz, 1156 High Street, Santa Cruz, CA 95060}
\author{Michael Bolte}
\affiliation{Department of Astronomy \& Astrophysics, University of California Santa Cruz, 1156 High Street, Santa Cruz, CA 95060}

\begin{abstract}
\label{abstract}
We report a method of identifying candidate low-metallicity blue compact dwarf galaxies (BCDs) from the Sloan Digital Sky Survey (SDSS) imaging data and present 3-m Lick Observatory and 10-m W.M. Keck Observatory optical spectroscopic observations of 94 new systems that have been discovered with this method. The candidate BCDs are selected from Data Release 12 (DR12) of SDSS based on their photometric colors and morphologies. Using the Kast spectrometer on the 3-m telescope, we confirm that the candidate low-metallicity BCDs are emission-line galaxies and we make metallicity estimates using the empirical $R$ and $S$ calibration methods. Follow-up observations on a subset of the lowest-metallicity systems are made at Keck using the Low Resolution Imaging Spectrometer (LRIS), which allow for a direct measurement of the oxygen abundance. We determine that 45 of the reported BCDs are low-metallicity candidates with 12\,+\,log(O/H)\,$\leq$\,7.65, including six systems which are either confirmed or projected to be among the lowest-metallicity galaxies known, at 1/30 of the solar oxygen abundance, or 12\,+\,log(O/H)\,$\sim$\,7.20.
\end{abstract}


\section{Introduction}
\label{intro}
The observed galaxy luminosity function (LF) shows that by number, low-luminosity galaxies dominate the total galaxy count of the Universe \citep{1976ApJ...203..297S}. The observed luminosity-metallicity ($L-Z$) relation \citep{1989ApJ...347..875S, 2001A&A...374..412P, 2009A&A...505...63G, 2012ApJ...754...98B}, which stems from the more fundamental mass-metallicity ($M-Z$) relation \citep{2004ApJ...613..898T, 2010MNRAS.408.2115M, 2015MNRAS.451.2251I}, shows that low-luminosity, low-mass galaxies are less chemically evolved than more massive galaxies, presumably due to less efficient star formation and higher metal loss during supernovae events and galactic-scale winds \citep{2009A&A...505...63G}.

The metallicity, $Z$, of a galaxy can be given in terms of the gas-phase oxygen abundance, denoted by 12\,+\,log(O/H). A galaxy is defined to be low metallicity if it has a gas-phase oxygen abundance 12+log(O/H) $\leq$ 7.65. This corresponds to $\lesssim$\,0.1\,$Z_{\odot}$ \citep{2000A&ARv..10....1K, 2007A&A...464..859P, 2010MNRAS.406.1238E}, where solar metallicity $Z_{\odot}$ is equivalent to an oxygen abundance of 12\,+\,log(O/H)\,=\,8.69 \citep{2009ARA&A..47..481A}. Despite the expected large population of low-luminosity galaxies from the LF, this low-mass, low metallicity regime is still relatively under-studied. As a result, observationally-derived properties such as the $L-Z$ and $M-Z$ relations are not well constrained at the low metallicity end. Progress towards identifying new metal-poor systems has been relatively slow due to their intrinsic low surface brightnesses that push on our current observational limits. Identifying these faint galaxies requires that they are relatively nearby, or that they contain bright O or B stars due to an episode of recent star formation. Because these galaxies are inefficient at forming stars, there is an additional caveat that these galaxies tend to be captured only during a brief stage of star formation, when ionized \HII\ regions are illuminated by the most massive stars.

%Despite the expected large population of low-luminosity galaxies from the LF, this low-mass, low metallicity regime is still relatively under-studied in the literature. Observationally derived properties such as the $L-Z$ and $M-Z$ relations are therefore not-well constrained at the low metallicity end. Citing the metallicity, $Z$, in terms of the gas phase oxygen abundance, denoted by 12+log(O/H), and adopting a solar metallicity, $Z_{\odot}$, of 12\,+\,log(O/H)\,=\,8.69 \citep{2009ARA&A..47..481A}, a galaxy is defined to be of low metallicity if it has a gas phase oxygen abundance  12+log(O/H) $\leq$ 7.65, corresponding to $\lesssim$\,0.1\,$Z_{\odot}$  \citep{2000A&ARv..10....1K, 2007A&A...464..859P, 2010MNRAS.406.1238E}. Progress towards identifying new metal-poor systems has been relatively slow due to their intrinsic low surface brightnesses that push on our observational limits. These faint galaxies elude detection unless they are relatively nearby, or if they contain bright O star due to an episode of recent star formation. Because these galaxies are also inefficient at forming stars, there is an additional caveat that these galaxies tend to be captured during a brief stage of star formation, when ionized \HII\ regions are illuminated by the most massive stars.

Observations of low-metallicity galaxies are important for a variety of studies, such as measurements of the primordial abundances \citep{1992MNRAS.255..325P, b9890a982907415fa4d3eb85829e9388, 2013AJ....146....3S, 2014MNRAS.445..778I, 2015JCAP...07..011A}, the formation and properties of the most metal-poor stars in primitive galaxies \citep{2005ApJS..161..240T}, and how these massive stars interacted with their surroundings \citep{2008Sci...319..174M, 2017AA...608A.119C}. Additionally, low-mass, low-metallicity systems are thought to be main contributors to the reionization of the Universe at high-redshifts, and local counterparts to these star-forming dwarf galaxies at high-redshifts are promising candidates for studies on leaking ionizing radiation from these systems and the effect on the surrounding intergalactic medium (IGM; \citealt{2015A&A...576A..83S, 2016Natur.529..178I, 2018MNRAS.474.4514I}). In our local Universe, studies of low-metallicity galaxies tend to focus on blue compact dwarf galaxies (BCDs; also referred to in the literature as extremely metal-poor galaxies, XMPs, or extremely metal-deficient galaxies, XMDs) because the presence of recent or actively forming massive stars within these galaxies ionize their surroundings, creating \HII\ regions from which emission lines can be easily detected. 
%In the local universe, blue compact dwarf galaxies (BCDs; also referred to in literature as extremely metal-poor galaxies, or XMPs), are ideal systems for observing due to the presence of massive stars within these galaxies that ionize their surroundings, creating \HII\ regions from which emission-lines can be easily detected.
%Studies of nearby, low metallicity galaxies tend to focus on blue compact dwarf galaxies (BCDs; also referred to in literature as extremely metal-poor galaxies, or XMPs) because the presence of current or actively forming massive stars within these galaxies ionize their surroundings, creating \HII\ regions from which emission-lines can easily be detected.

Hydrogen and helium recombination emission line ratios observed in these BCDs, combined with direct measurements of their gas-phase oxygen abundance, allow for constraints on the primordial helium abundance produced during Big Bang Nucleosynthesis (BBN; \citealt{2007ARNPS..57..463S, 2016RvMP...88a5004C}). Observational measurements of the primordial helium abundance from galaxies provide an important cross-test on the standard cosmological model and its parameters as obtained by the Wilkinson Microwave Anisotropy Probe (WMAP; \citealt{2013ApJS..208...19H}) and \textit{Planck} \citep{2016AA...594A..13P}. A recent study by \citet{2014MNRAS.445..778I} using low-metallicity \HII\ regions to observationally constrain the primordial helium abundance indicated a slight deviation from the Standard Model, suggesting tentative evidence of new physics at the time of BBN. However, analysis on the same dataset in a follow-up work by \citet{2015JCAP...07..011A} found a different value of the primordial helium abundance, one that is in agreement with that of the Standard Model (see also \citealt{2017RMxAC..49..181P}). The disagreement between the most recent determinations of the primordial helium abundance suggests that underlying systematics may not be fully accounted for. Currently, the number of low-metallicity systems available for primordial abundance measurements is limited, especially in the \textit{lowest} metallicity regime.  Increasing the number of metal-poor galaxies in the lowest metallicity regime to further our understanding of the primordial helium abundance is a key goal of our survey.

BCDs contain a significant fraction of gas and to be experiencing a recent burst of star formation ($\lesssim$ 500 Myr ago). The proximity of local BCDs allow for detailed studies of their stellar and gas content and the physical conditions of dwarf galaxies. These physical properties characterize the conditions under which the first stars might have formed and the various processes that trigger and suppress star formation in dwarfs \citep{2004ApJ...613..898T, 2016Natur.535..523F}. The first stars are believed to be a massive generation of stars that synthesized then enriched their host minihalos with the first chemical elements heavier than lithium \citep{2002ApJ...564...23B}. Detailed studies of BCDs allow us to better understand the physics of how early galaxies might have been enriched and affected by the first generation of massive stars \citep{2001ApJ...555...92M, 2003ApJ...588...18F, 2008ApJ...685...40W}. Despite their burst of recent or on-going star formation and low metallicities that may suggest these systems to be young galaxies, well-studied dwarf galaxies such as Leo P \citep{2015ApJ...812..158M} and I Zwicky 18 \citep{2007ApJ...667L.151A} have been found to be at least $\sim$10 Gyr old, evidenced by the detection of an RR Lyrae or red giant branch (RGB) population. Local BCDs thus provide insight on the star formation histories (SFH) of dwarf galaxies, which can constrain the initial mass function (IMF) in the low metallicity regime, which is currently not well established, but thought to be dominated by high-mass stars, in contrast to the present day stellar IMF \citep{2002ApJ...564...23B, 2012MNRAS.422.2246M, 2013ApJ...766..103D}.

%Understanding how the metal content evolves with the mass of a galaxy is crucial in understanding the physical processes, such as inflows and outflows, affecting these systems and their surroundings. In particular, 
Low-mass, star-forming galaxies are thought to contribute significantly to the reionization of the Universe by redshift z$\sim$6 \citep{2009ApJ...693..984W, 2016Natur.529..178I} due to leaking ionizing radiation from the galaxies. Although observations of the population of low-mass, high-redshift systems are limited, it has been found that low-redshift compact star-forming galaxies follow similar $M-Z$ and $L-Z$  relations as higher-redshift star-forming galaxies \citep{2015MNRAS.451.2251I}. Local BCDs are therefore important proxies for studies of the higher redshift Universe, particularly in constraining the faint end slope of the $M-Z$ relation and in understanding how radiation and material from low-mass systems are redistributed to their environments. These studies can then inform models on the nature and timing of how the IGM was reionized during the epoch of reionization \citep{2013MNRAS.428.1366J}. Additionally, understanding the mass loss in low-mass galaxies allows for studies on the metal retention of dwarf galaxies and subsequently, on the chemical evolution of this population of galaxies.

It is necessary, however, to increase the number of the \textit{lowest} metallicity BCDs to make better primordial helium abundances measurements, study the low-mass and low-luminosity regimes that these metal-deficient galaxies define, and better understand the physical and chemical evolution of these systems. Only a handful of systems are currently known with metallicities of $\lesssim$\,0.03\,$Z_{\odot}$, or 12\,+\,log(O/H)\,$\lesssim$\,7.15. Efforts toward identifying new low-metallicity systems have typically focused on discoveries through emission-line galaxy surveys \citep{2012A&A...546A.122I, 2017RAA....17...41G, 2017A&A...599A..65G, 2017ApJ...847...38Y}, with limited results on identifying new systems that push on the lowest metallicity regime. Although the well-known higher-luminosity, metal-poor systems I Zwicky 18 \citep{1966ApJ...143..192Z}, SBS-0335-052 \citep{1990Natur.343..238I}, and DDO68 \citep{2005A&A...443...91P} have been known for several decades, progress in discovering the most metal-poor systems has been slow. Leo P \citep{2013AJ....146...15G, 2013AJ....146....3S} and AGC198691 \citep{2016ApJ...822..108H}, both having been discovered through the \HI\ 21\,cm Arecibo Legacy Fast ALFA (ALFALFA; \citealt{2005AJ....130.2598G, 2011AJ....142..170H}) survey, the Little Cub \citep{2017ApJ...845L..22H}, and J0811$+$4730 \citep{2018MNRAS.473.1956I}, discovered through Sloan Digital Sky Survey (SDSS) photometry and spectroscopy respectively, are the recent exceptions. James et al. (\citeyear{2015MNRAS.448.2687J, 2017MNRAS.465.3977J}) conducted a photometric search for low metallicity objects and obtained follow-up spectroscopy on a subset of their sample. Using this photometric method, \citeauthor{2017MNRAS.465.3977J} found a higher success rate in identifying low metallicity systems, with $\sim$20\% of their observed sample being $\leq$\,0.1\,$Z_{\odot}$, though none of their sample had gas phase oxygen abundances of 12\,+\,log(O/H)\,$\lesssim$\,7.45.

Eliminating the need for existing spectroscopic information can be a method of efficiently increasing the known population of BCDs, particularly at the lowest metallicities, since this allows a targeted spectroscopic campaign of the lowest-metallicity galaxies based on photometry alone. In Section \ref{candidate_selection}, we describe a new photometric query designed to identify new metal-poor BCDs in our local Universe using only photometric data from SDSS. Observations of a subset of candidate BCDs, along with data reduction procedures are described in Section \ref{ODR}. We discuss emission line measurements, present gas phase oxygen abundances, and derive metallicities of 94 new systems in Section \ref{AD}, and calculate the distance, H$\alpha$ luminosity, star formation rate, and stellar mass to each system. In Section \ref{BCDsXMPs}, we discuss our sample of BCDs in the context of the population of metal-poor systems as a whole and consider other photometric surveys that offer a means of discovering BCDs, both locally as in SDSS, as well as pushing towards higher redshift. Our findings are summarized in Section \ref{C}.

\section{Candidate Selection}
\label{candidate_selection}
\subsection{Photometric Selection}
%JXP -- This sub-section needs a supporting figure.
To identify candidate low-metallicity BCDs, we conducted a query for objects in SDSS Data Release 12 (DR12) with photometric colors similar to those of currently known low-metallicity systems, including Leo P and I Zwicky 18. This color selection criteria will be biased towards finding BCDs at low redshift, corresponding to the colors of Leo P and I Zwicky 18; our color selection criteria does not account for the redshift evolution of BCD colors, which is the goal of a future work (Tirimba et al. in prep.). We require that the objects lie outside of the galactic plane, i.e., have Galactic latitudes $b > +25$\,deg 
and $b < -25$\,deg, 
have $r$-band magnitudes $r \leq 21.5$, and fall within the following
color cuts:
\begin{gather*} 
0.2 \leq u-g \leq 0.6 \\
-0.2 \leq g-r \leq 0.2 \\
-0.7 \leq r-i \leq -0.1 \\
-0.4 - 2 \, z_\textnormal{error} \leq i-z \leq 0.1
\end{gather*}
Here, the magnitudes are given as inverse hyperbolic sine magnitudes (``asinh" magnitudes; \citealt{1999AJ....118.1406L}). The $2z_{error}$ term ensures a 2$\sigma$ lower bound on objects with a poorly constrained $z$-band magnitude. We also require the SDSS $g$-band fiber magnitude to be less than the $z$-band fiber magnitude to exclude \HII\ regions in redder galaxies from the query results. Finally, we require that the objects be extended, i.e., classified as a Galaxy in SDSS. This query returned a total of 2505 candidate objects. Our full query is presented in Appendix \ref{CasJobsquery}.

\begin{comment}
SELECT P.ObjID, P.ra, P.dec, P.u, P.g, P.r, P.i, P.z into mydb.MyTable from Galaxy P
WHERE
  ( P.u - P.g > 0.2)
  and ( P.u - P.g < 0.60)
  and ( P.g - P.r > -0.2 )
  and ( P.g - P.r <  0.2 )
  and ( P.r - P.i < -0.1 )
  and ( P.r - P.i > -0.7 )
  and ( P.i - P.z < 0.1 )
  and ( P.i - P.z > -0.4 - 2*P.err_z )
  and ( P.r < 21.5)
  and (( P.b < -25.0) or ( P.b > 25.0) )
  and (P.fiberMag_g<P.fiberMag_z)
\end{comment}

\subsection{Morphological Selection}
To create a list of candidate objects best fit for observation, we individually examined the SDSS imaging of the 2505 objects from the photometric query. This procedure eliminated objects misclassified as individual galaxies, such as stars or star-forming regions located in the spiral arms of larger galaxies, and predisposes our candidate list towards systems in isolated environments. We also eliminated objects with existing SDSS spectra.
%[MB: not obvious why we do this as you can immediately sort identify emission-line systems] . We had mentioned above that finding the lowest metallicity systems from emission-line surveys typically yields few results
The remaining candidate galaxies that appeared to have a bright knot surrounded by a dimmer, more diffuse region were chosen as ideal systems for follow up spectroscopic observations, with the assumption that active star-forming \HII\ regions would appear as bright `knots' in SDSS imaging and would be the most likely to yield easily detectable emission lines. The surrounding diffuse region is assumed to be indicative of the remaining stellar population in the system. This selection criteria was not quantified, but is similar to the ``single knot'' morphological description as presented in \cite{2011ApJ...743...77M}.

\begin{figure}
\includegraphics[width=\linewidth]{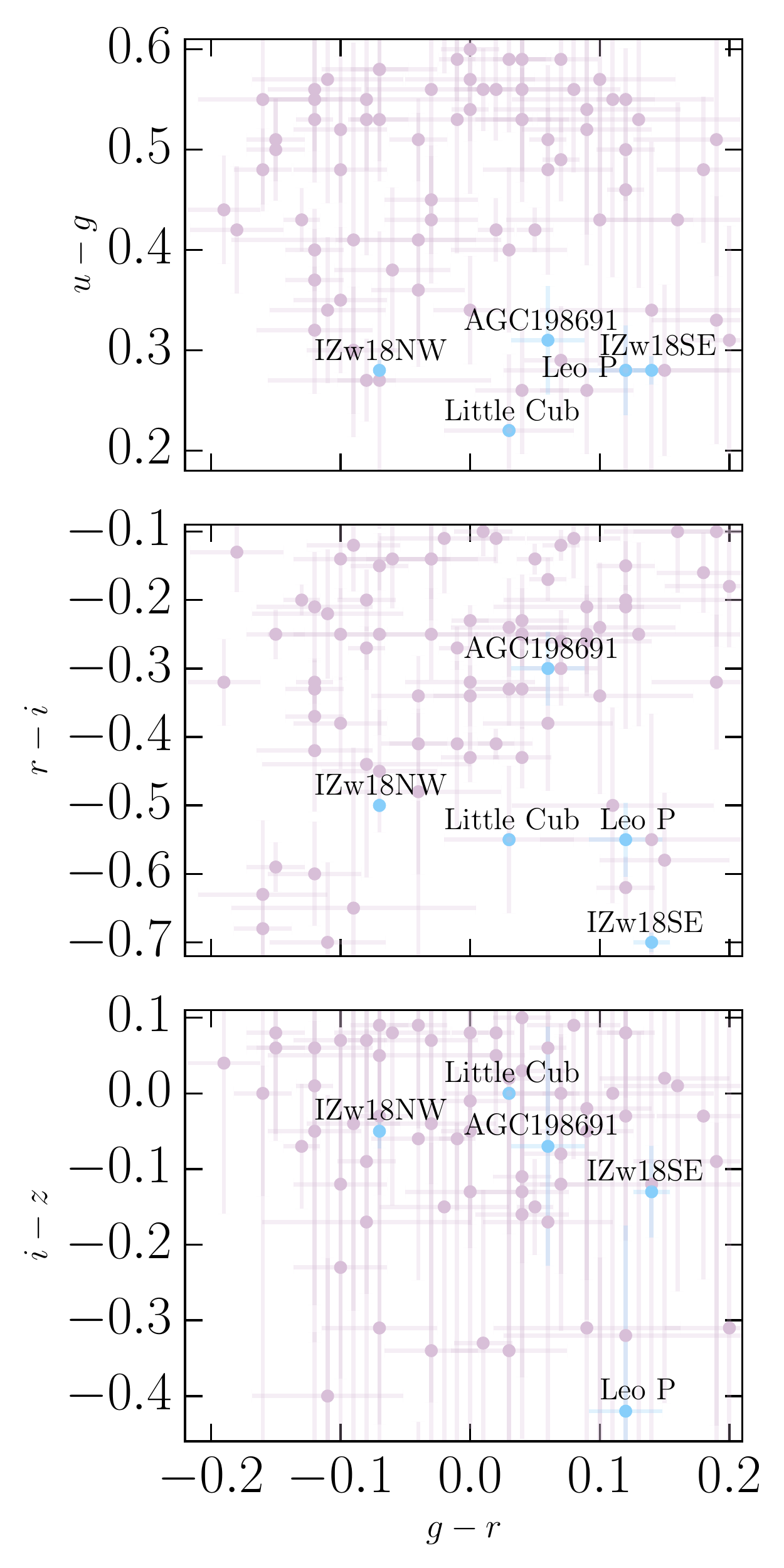}
\caption{Our SDSS \textit{g-r} color selection criteria versus \textit{u-g}, \textit{r-i}, and \textit{i-z} color cuts in the upper, middle, and lower panels, respectively. The purple points represent the location in color-color space of candidate BCDs selected for observing. The blue points show the location of the known, extremely metal-poor systems such as Leo P and I Zwicky 18 (both the northwest and southeast components), in the same color-color space. Error bars on the colors are shown. We note that Leo P and I Zwicky 18 were known systems prior to this survey and helped define our color-color search criteria, whereas AGC198691 and the Little Cub were identified as a result of the query. The lowest metallicity systems appear to cluster around $u-g\,\sim\,0.27$ and $i-z\,\sim\,-0.06$, with the exception of Leo P in the latter.}
\label{fig:colorcuts}
\end{figure}

\begin{figure*}
\centering
\includegraphics[width=\linewidth]{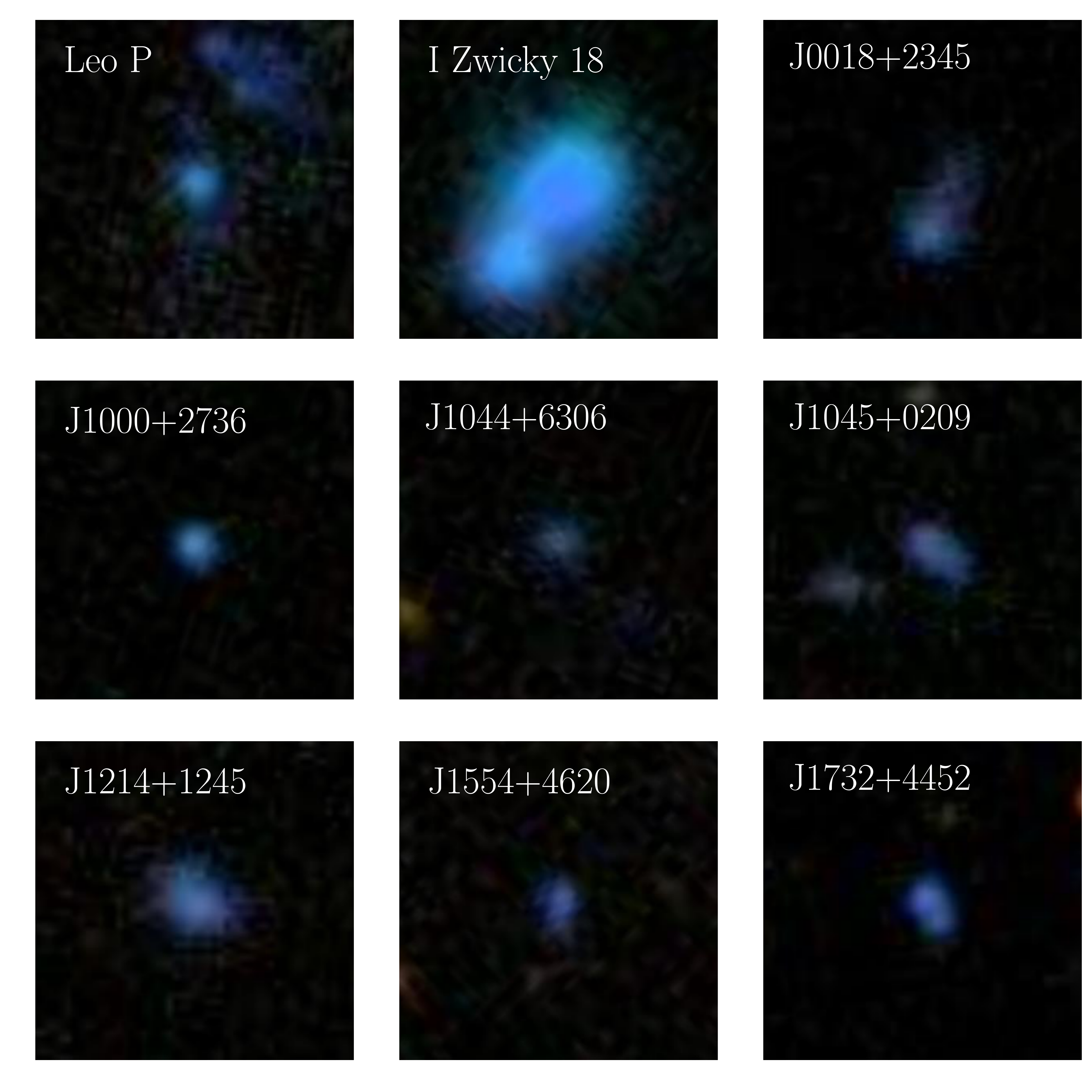}
\caption{SDSS imaging of Leo P and I Zwicky 18, two of the most metal-poor \HII\ regions currently known, are shown in the left and middle panels of the upper row. The remaining panels show SDSS imaging of seven \HII\ regions selected for observing via our photometric method and predicted to be of low metallicity. Spectra corresponding to the new systems are shown in Fig \ref{fig:spectrumpanel}. The images are shown on the same angular scale of 15$''$ on a side, with north up and east to the left.}
\label{fig:SDSSimg}
\end{figure*}

Our morphological selection criteria condensed the candidate list down to 236 objects. To date, we have observed 154 of the selected candidate BCDs, with the 154 objects having RAs best fit for our scheduled observing nights. The candidate systems we have targeted so far are shown in SDSS color-color space in Figure \ref{fig:colorcuts}. A subset of these BCDs in SDSS imaging are shown in Figure \ref{fig:SDSSimg}.

% Following two figures made in: ~/old_science_frames/plot_image_spectrum_panels

\section{Observations and Data Reduction}
\label{ODR}
To confirm the candidate BCDs as galaxies and identify the \textit{lowest} metallicity systems, we require spectroscopic observations for preliminary estimates of the oxygen abundance. We use the $R$ and $S$ calibration methods presented by \citet{2016MNRAS.457.3678P},
which compare the strengths of the metal [\OII], [\OIII], [\NII], and [\SII] emission lines to the H Balmer emission lines, and allow for an approximate measurement of the metallicity of the system. Specifically, the emission lines targeted with our survey include: the forbidden [\OII] doublet at $\lambda\lambda$3727,3729\AA $ $,  H$\beta$ emission at $\lambda$4861\AA $ $, a forbidden [\OIII] doublet at $\lambda\lambda$4959,5007\AA $ $, H$\alpha$ emission at $\lambda$6563\AA $ $, a forbidden [\NII] doublet at $\lambda\lambda$6548,6583\AA $ $, and a forbidden [\SII] doublet at $\lambda\lambda$6717,6731\AA $ $. Detecting these lines are the goal of our initial observations, which were mostly made using the Shane 3-m telescope at Lick Observatory.

For observations made at Keck Observatory, where we can achieve a higher signal-to-noise (S/N) ratio and therefore a greater sensitivity to weak emission lines, 
we aim to detect the temperature sensitive [\OIII] $\lambda$4363\AA\,line for a direct measurement of the oxygen abundance. Additionally, with the Keck observations, we aim to detect at least five optical \HeI\ emission lines to reliably determine the physical state of the \HII\ regions, which is necessary for primordial helium studies.
%[MB: He lines are only one aspect of determining physical state. Perhaps want to explain more fully how five He lines are crucial ]

\begin{deluxetable*}{ccccccccc}[ht]
\tablewidth{12pt}
\tablecaption{Observational and measured properties of our BCD sample, observed at Lick and Keck Observatory} 
\tablehead{ 
\colhead{Target Name} & \colhead{RA} & \colhead{DEC} & \colhead{Observations} & \colhead{\textit{z}} & \colhead{Distance} & \colhead{$m_{g}$} & \colhead{$u-g$} & \colhead{12\,+\,log(O/H)} \\
\colhead{} & \colhead{(J2000)} & \colhead{(J2000)} & \colhead{} & \colhead{} & \colhead{(Mpc)} & \colhead{} & \colhead{} & \colhead{}
}
\startdata 
J0000$+$3052A & 00$^{\textnormal{h}}$00$^{\textnormal{m}}$31$^{\textnormal{s}}$.45 & $+$30$^{\circ}$52$'$09.30$''$ & Keck+LRIS & 0.0151 & 67.6 & 19.84\,$\pm$\,0.02 & 0.65 & 7.72\,$\pm$\,0.01 \\
J0000$+$3052B & 00$^{\textnormal{h}}$00$^{\textnormal{m}}$32$^{\textnormal{s}}$.31 & $+$30$^{\circ}$52$'$16.62$''$ & Keck+LRIS & 0.0153 & 68.5 & 19.55\,$\pm$\,0.02 & 0.43 & 7.63\,$\pm$\,0.02 \\
J0003$+$3339 & 00$^{\textnormal{h}}$03$^{\textnormal{m}}$51$^{\textnormal{s}}$.08 & $+$33$^{\circ}$39$'$29.63$''$ & Shane+Kast & 0.0211 & 94.7 & 19.57\,$\pm$\,0.03 & 0.33 & 7.91\,$\pm$\,0.03 \\
J0018$+$2345 & 00$^{\textnormal{h}}$18$^{\textnormal{m}}$59$^{\textnormal{s}}$.32 & $+$23$^{\circ}$45$'$40.32$''$ & Shane+Kast & 0.0154 & 68.8 & 19.24\,$\pm$\,0.02 & 0.26 & 7.18\,$\pm$\,0.03 \\
J0033--0934 & 00$^{\textnormal{h}}$33$^{\textnormal{m}}$55$^{\textnormal{s}}$.79 & --09$^{\circ}$34$'$32.20$''$ & Shane+Kast & 0.0121 & 54.2 & 17.97\,$\pm$\,0.01 & 0.55 & 7.84\,$\pm$\,0.23 \\
J0035--0448 & 00$^{\textnormal{h}}$35$^{\textnormal{m}}$39$^{\textnormal{s}}$.64 & --04$^{\circ}$48$'$40.93$''$ & Shane+Kast & 0.0169 & 75.9 & 19.58\,$\pm$\,0.02 & 0.53 & 7.64\,$\pm$\,0.02 \\
J0039$+$0120 & 00$^{\textnormal{h}}$39$^{\textnormal{m}}$30$^{\textnormal{s}}$.30 & $+$01$^{\circ}$20$'$21.61$''$ & Shane+Kast & 0.0147 & 66.0 & 19.88\,$\pm$\,0.03 & 0.52 & 7.78\,$\pm$\,0.03 \\
J0048$+$3159 & 00$^{\textnormal{h}}$48$^{\textnormal{m}}$55$^{\textnormal{s}}$.31 & $+$31$^{\circ}$59$'$02.05$''$ & Shane+Kast & 0.0153 & 68.5 & 19.05\,$\pm$\,0.03 & 0.31 & 8.45\,$\pm$\,0.02 \\
J0105$+$1243 & 01$^{\textnormal{h}}$05$^{\textnormal{m}}$24$^{\textnormal{s}}$.95 & $+$12$^{\circ}$43$'$38.71$''$ & Shane+Kast & 0.0142 & 63.4 & 19.77\,$\pm$\,0.04 & 0.43 & 7.64\,$\pm$\,0.05 \\
J0118$+$3512 & 01$^{\textnormal{h}}$18$^{\textnormal{m}}$40$^{\textnormal{s}}$.00 & $+$35$^{\circ}$12$'$57.0$''$ & Keck+LRIS & 0.0165 & 73.9 & 19.23\,$\pm$\,0.02 & 0.27 & 7.58\,$\pm$\,0.01 \\
\enddata 
\tablecomments{Distances reported in this table are luminosity distances, assuming a \textit{Planck} cosmology \protect \citep{2016AA...594A..13P}. All metallicity estimates for systems observed on Shane+Kast are determined using the $R$ and $S$ calibration methods, with the reported metallicity being the average of the $R$ and $S$ methods. All BCDs observed using Keck+LRIS have direct metallicity calculations, except for J0743$+$4807, J0812$+$4836, J0834$+$5905. In these cases, we do not significantly detect the [\OIII] $\lambda$4363\AA\,line and adopt an upper limit to the [\OIII] $\lambda$4363\AA\,emission line flux equivalent to 3 times the error in the measured line flux at that wavelength. This results in a lower limit on their metallicities. The asterisk (*) in the Observations column indicates that observations were made first using Lick+Kast, with follow-up made using Keck+LRIS. For such systems, the derived values reported here are measurements from the Keck+LRIS observations. Values for the full sample of BCDs are available online.} 
\label{tab:observations}
\end{deluxetable*}

\subsection{Lick Observations}
Spectroscopic observations of 135 candidate BCDs were made using the Kast spectrograph on the Shane 3-m telescope at Lick Observatory over 22 nights during semesters 2015B, 2016A, and 2016B. 85 of the observed candidates yielded emission line detections, and 78 of the 85 have confident emission line measurements reported here. 
% 78 from Kast + 29 from LRIS - 13 re-observed on LRIS = 94

The Kast spectrograph has separate blue and red channels, which our observational setup utilized simultaneously. Observations made prior to 6 October 2016 were obtained using the d55 dichroic, with the Fairchild 2k $\times$ 2k CCD detector on the blue side and the Reticon 400\,$\times$\,1200 CCD detector on the red side. Thereafter, the d57 dichroic was used, along with a Hamamatsu 1024 $\times$ 4096 CCD detector on the red side. The pixel scale on the Reticon is 0.78$''$\,per pixel, and 0.43$''$\,per pixel on the Fairchild and Hamamatsu devices. On the blue side, the 600/4310 grism with a dispersion of 1.02~\AA\,pix$^{-1}$ was used, while on the red side, the 1200/5000 grating with a dispersion of 0.65~\AA\,pix$^{-1}$ was used. 
This instrument setup covers $\sim$3300--5500~\AA\,and  $\sim$5800--7300~\AA, with instrument full-width at half maximum (FWHM) resolutions of 6.4~\AA\,and 2.7~\AA,
%JXP -- Can you give the FWHM in km/s?
in the blue and red, respectively. This allows for sufficient coverage and spectral resolution of all emission lines of interest. Specifically, we are able to resolve the [\NII] doublet from H$\alpha$. However, we note that the [\OII] doublet is not resolved with this setup.

All targets were observed using a 2$''$ slit and at the approximate
parallactic angle to mitigate the effects of atmospheric diffraction. Total exposure times range from 3\,$\times$\,1200~s to 3\,$\times$\,1800~s for our objects. Spectrophotometric standard stars were observed at the beginning and end of each night for flux calibration. Spectra of the Hg-Cd and He arc lamps on the blue side and the Ne arc lamp on the red side were obtained at the beginning of each night for wavelength calibrations.
%[MB: presumably it was not always photometeric, but the clouds are gray and the relative flux as a function of wavelength is determined and applied. Is that right? If so, might want to mention it]
Bias frames and dome flats were also obtained to correct for the detector bias level and pixel-to-pixel variations, respectively. The RA and DEC, measured redshift, estimated distance, $g$-band magnitude, $u$-$g$ color, and gas phase oxygen abundance of a selection of observed and confirmed emission-line systems are reported in Table \ref{tab:observations}; the full sample of observed systems is available online.

\subsection{Keck Observations}
Spectroscopic observations of 29 candidate BCDs were made using the Low Resolution Imaging Spectrometer (LRIS) at the W.M. Keck Observatory over a three night program during semesters 2015B and 2016A. Thirteen observations made using LRIS were emission-line galaxies previously observed using the Kast spectrograph, with the remaining objects having only LRIS data. Similar to the Kast spectrograph, LRIS has separate blue and red channels. Our setup utilized the 600/4000 grism on the blue side , which provides a dispersion of 0.63~\AA\,pix$^{-1}$. On the red side, the 600/7500 grating provides a dispersion of 0.8~\AA\,pix$^{-1}$. Using the D560 dichroic, the full wavelength coverage achieved with this instrument setup is $\sim$3200--8600\AA, with the blue side covering $\sim$3200--5600\AA\,and the red side covering $\sim$5400--8600\AA. The blue and red channels have FWHM resolutions of 2.6~\AA\,and 3.1~\AA\,respectively. We note that while the separate blue and red arms overlap in wavelength coverage, data near the region of overlap can be compromised due to the dichroic.
%[MB: do the red and blue side wavelengths really overlap?]

All targets were observed using a $0.7^{\prime\prime}$ slit using the atmospheric dispersion corrector (ADC) on LRIS %at parallactic angle,
for total exposure times ranging from 3\,$\times$\,1200~s to 3\,$\times$\,1800~s. Bias frames and dome flats were obtained at the beginning of the night, along with spectra of the Hg, Cd, and Zn arc lamps on the blue side and Ne, Ar, Kr arc lamps and red side for wavelength calibration. Photometric standard stars were observed at the beginning and end of each night for flux calibration. Observed and derived physical properties of a sample of BCDs observed using Keck+LRIS are reported in Table \ref{tab:observations}. For systems observed both at Lick and Keck, we present properties derived from observations made using Keck+LRIS and note the systems with an asterisk. The full sample of observed systems is available online.

% 0.135"/pix on LRISb, LRISr
% 0.43"/pix, 0.78"/pix on KASTb, KASTr

\subsection{Data Reduction}
\begin{figure*}
\centering
% made in ~/BCDs/old_science_frames/plot_image_spectrum_panels.ipynb
\includegraphics[width=0.85\textwidth]{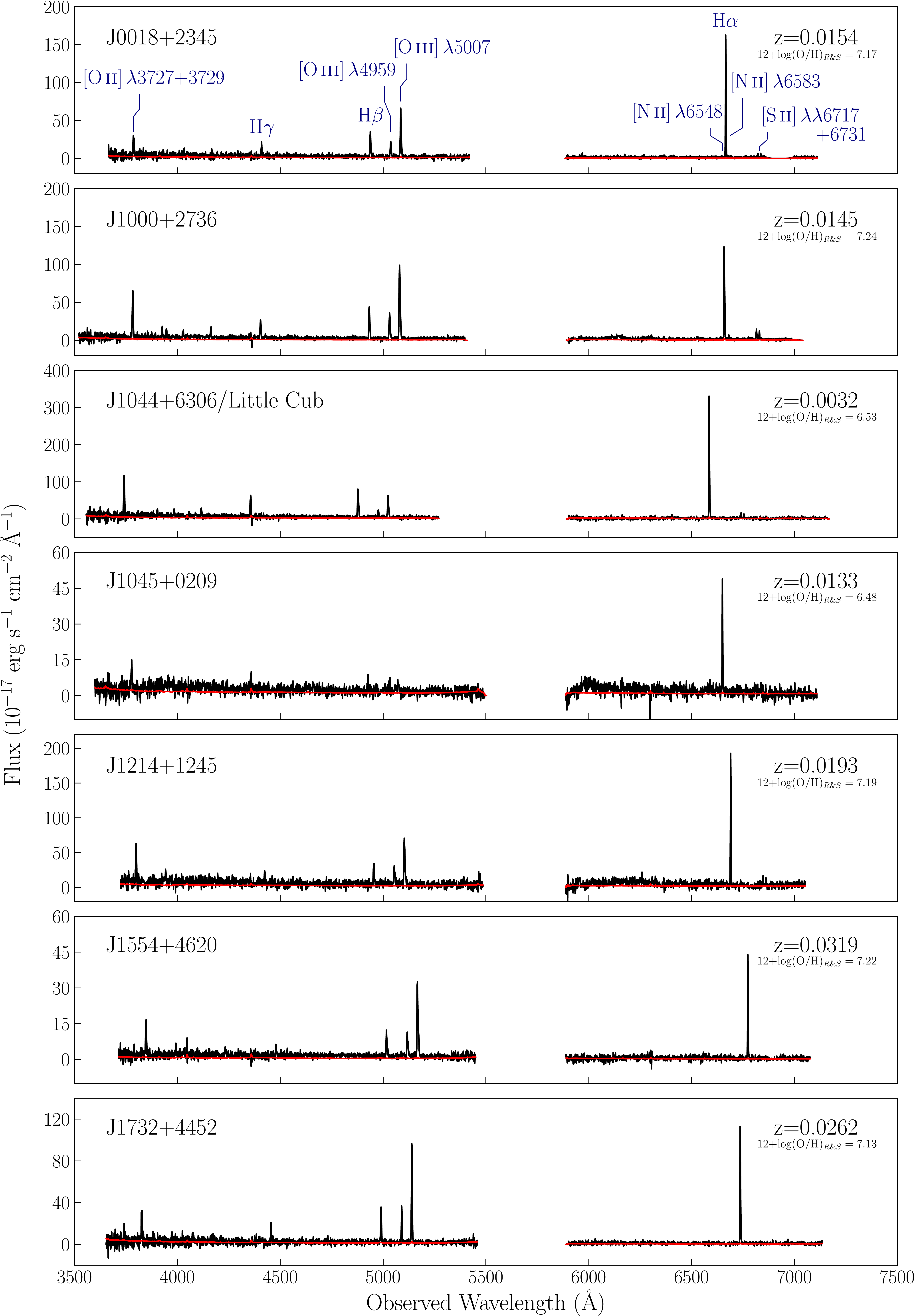}
\caption{Discovery spectra (shown in black) obtained using the Shane 3-m telescope at Lick Observatory of seven \HII\ regions in our sample that are predicted to have the lowest metallicities via the $R$ and $S$ calibration methods. The error spectra are shown in red. Emission lines of interest for the $R$ and $S$ calibration methods are labeled in the first panel. The gap between $\sim$5400--5900\,\AA\,in all panels is due to the d55 dichroic used during our observations on the Kast spectrograph. We note that the object named J1044$+$6306 is the Little Cub, as presented in \citet{2017ApJ...845L..22H} and is henceforth referred to as the Little Cub.}
%These seven systems have some of the lowest predicted metallicities based on these initial 3-m observations, which helped prioritize objects for follow-up Keck+LRIS spectroscopy. 
\label{fig:spectrumpanel}
\end{figure*}

The two-dimensional raw images were individually bias subtracted, flat-field corrected, cleaned for cosmic rays, sky-subtracted, extracted, wavelength calibrated, and flux calibrated, using PYPIT, a Python based spectroscopic data reduction package.\footnote{\textsc{pypit} is available from: \url{https://github.com/PYPIT/PYPIT}} PYPIT applies a boxcar extraction to extract a one-dimensional (1D) spectrum of the object. Multiple exposures on a single candidate BCD were combined by weighting each frame by the inverse variance at each pixel.
%[MB: which variance? At a particular emission line?]
The reduced and combined spectra of seven BCDs observed at Lick Observatory are shown in Figure \ref{fig:spectrumpanel}.

\section{Analysis and Discussion}
\label{AD}
\subsection{Emission Line Measurements}
Emission line fluxes were measured using the Absorption LIne Software (ALIS\footnote{\textsc{ALIS} is available from: \url{https://github.com/rcooke-ast/ALIS/}}; see \citealt{2014ApJ...781...31C} for details of the software), which performs spectral line fitting using $\chi^{2}$ minimization. For integrated flux measurements, each emission line is fit with a Gaussian model simultaneously with the surrounding continuum, which is modeled with a first order Legendre polynomial. In this procedure, the error in the continuum measurement is folded into the integrated flux errors. We assume that the full width at half maximum (FWHM) of all emission lines are set by instrumental broadening, and therefore all emission lines have the same FWHM. %that the intrinsic width of the emission lines is much smaller than this. %the typical internal velocities of \HII\ regions, chosen to be 30 km/s, convolved with the instrument resolution of 150 km/s. %All emission lines are tied to have the same redshift, and the [\OIII] $\lambda\lambda$4959,5007\AA $ $ and [\NII] $\lambda\lambda$6548,6583\AA $ $ doublets are set to have emission lines measurements locked in a 3:1 ratio.
The integrated flux measurements of our observed systems are available online.

The measured emission line fluxes are corrected for reddening and underlying stellar absorption using the $\chi^{2}$ minimization approach described below and found in Appendix A of \citet{b9890a982907415fa4d3eb85829e9388}:\footnote{We note that our numerator in Equation \ref{equation:chi_squared} differs slightly from that given in Appendix A of \citet{b9890a982907415fa4d3eb85829e9388} due to a typographical error in the original work (E. Skillman, private communication).}
\begin{equation}
\chi^{2}\,=\,\sum_{\lambda}\frac{\big(X_{R}(\lambda)\,-\,X_{T}(\lambda)\big)^{2}}{\sigma^{2}_{X_{R}}(\lambda)} \\
\label{equation:chi_squared}
\end{equation}
where
\begin{equation}
X_{R}(\lambda)\,=\,\frac{I(\lambda)}{I(H\beta)}\,=\,\frac{X_{A}(\lambda)}{X_{A}(H\beta)\,10^{f(\lambda)c(H\beta)}} \\
\end{equation}
\begin{equation}
X_{A}(\lambda)\,=\,F(\lambda)\,\Big(\frac{W(\lambda)\,+\,a_{\textnormal{\HI}}}{W(\lambda)}\Big)
\end{equation}
Here, $X_{T}(\lambda)$ is the theoretical value of the Balmer line ratio at wavelength $\lambda$ of consideration to H$\beta$, $f$($\lambda$) is the reddening function, normalized at H$\beta$, c(H$\beta$) is the reddening, $W(\lambda)$ is the equivalent width of the line, and $a_{\textnormal{H}\textsc{i}}$ is the equivalent width of the underlying stellar absorption at H$\beta$, both given in Angstroms. Minimizing the value of $\chi^{2}$ allows for the determination of the best values of c(H$\beta$) and $a_{\textnormal{H}\textsc{i}}$.

We note that the underlying stellar absorption is wavelength dependent. While we report the value of $a_{\textnormal{H}\textsc{i}}$ at H$\beta$, the best solution for the $\chi^{2}$ minimization is the parameter that fits all Balmer line ratios used in the analysis, where the correction to each Balmer line ratio is applied as $a_{\textnormal{H}\textsc{i}}$ times a multiplicative coefficient that accounts for the wavelength dependence of underlying stellar absorption. The multiplicative coefficients we applied are given in Equation 5.1 of \citet{2010JCAP...05..003A} and we refer readers to Section 5 of \citet{2010JCAP...05..003A} for a more detailed discussion on the wavelength dependence of underlying stellar absorption.

There is some uncertainty in the relative flux calibration across the separate blue and red channels on Kast and on LRIS;  H$\alpha$ is therefore not included in this calculation. Instead, we rely on all detected higher order Balmer lines when solving for the reddening and underlying stellar absorption. We include H$\beta$ through H9 in this calculation, and exclude H$\epsilon$ and H8 due to blends with [\NeIII] and \HeI, respectively. We note that the uncertainty in flux scales across the separate channels does not affect direct metallicity measurements, since all relevant emission lines for direct measurements fall on the blue detector. 

Throughout the procedure, we assume Balmer line ratios corresponding to a $T_{e}$\,=\,10,000~K gas for our Kast observations, and Balmer line ratios for measured temperatures are adopted for LRIS observations. The underlying stellar absorption in our systems range from $\lesssim$\,1\,\AA\,--\,4.5\,\AA, and the amount of reddening ranges from c(H$\beta$)\,$\sim$\,0.001\,--\,0.5. The measured emission line intensities for a few systems are shown in Table \ref{tab:flux}; the emission line intensities for our full sample of BCDs are available online.

\begin{deluxetable*}{cccccc}
\tablewidth{0.99\textwidth}
\tablecaption{Measured Emission Line Intensities for a sample of observed BCDs} 
\tablehead{
\colhead{} & \colhead{} & \colhead{} & \colhead{Target Name} & \colhead{} & \colhead{}}
\startdata 
Ion & J0000+3052A & J0000$+$3052B & J0003$+$3339 & J0018$+$2345 & J0033$-$0934 \\
\hline
{[O\,\textsc{ii}]\,$\lambda$3727+3729} & 0.6249\,$\pm$\,0.0064 & 1.413\,$\pm$\,0.011 & 1.532\,$\pm$\,0.040 & 0.799\,$\pm$\,0.044 & 1.378\,$\pm$\,0.052 \\
{H11\,$\lambda$3771} & 0.0785\,$\pm$\,0.0030 & \nodata & \nodata & \nodata & \nodata \\
{H10\,$\lambda$3798} & 0.0256\,$\pm$\,0.0048 & \nodata & \nodata & \nodata & \nodata \\
{H9\,$\lambda$3835} & 0.1187\,$\pm$\,0.0048 & 0.1140\,$\pm$\,0.0082 & 0.120\,$\pm$\,0.022 & 0.088\,$\pm$\,0.035 & 0.109\,$\pm$\,0.032 \\
{[Ne\,\textsc{iii}]\,$\lambda$3868} & 0.2528\,$\pm$\,0.0046 & 0.2025\,$\pm$\,0.0042 & 0.372\,$\pm$\,0.022 & 0.157\,$\pm$\,0.030 & 0.099\,$\pm$\,0.050 \\
{H8+He\,\textsc{i}\,$\lambda$3889} & 0.2006\,$\pm$\,0.0053 & 0.2202\,$\pm$\,0.0088 & 0.260\,$\pm$\,0.020 & 0.204\,$\pm$\,0.032 & 0.196\,$\pm$\,0.039 \\
{H$\epsilon$+[Ne\,\textsc{iii}]\,$\lambda$3968} & 0.2120\,$\pm$\,0.0055 & 0.2093\,$\pm$\,0.0088 & 0.172\,$\pm$\,0.022 & 0.228\,$\pm$\,0.034 & 0.120\,$\pm$\,0.048 \\
{H$\delta$\,$\lambda$4101} & 0.2518\,$\pm$\,0.0051 & 0.2681\,$\pm$\,0.0080 & 0.255\,$\pm$\,0.018 & 0.262\,$\pm$\,0.031 & 0.238\,$\pm$\,0.041 \\
{H$\gamma$\,$\lambda$4340} & 0.4289\,$\pm$\,0.0052 & 0.4437\,$\pm$\,0.0074 & 0.468\,$\pm$\,0.022 & 0.458\,$\pm$\,0.030 & 0.514\,$\pm$\,0.046 \\
{[O\,\textsc{iii}]\,$\lambda$4363} & 0.0861\,$\pm$\,0.0023 & 0.0518\,$\pm$\,0.0025 & \nodata & \nodata & \nodata \\
{He\,\textsc{i}\,$\lambda$4472} & 0.0309\,$\pm$\,0.0020 & 0.0244\,$\pm$\,0.0024 & \nodata & \nodata & \nodata \\
{H$\beta$\,$\lambda$4861} & 1.0000\,$\pm$\,0.0049 & 1.0000\,$\pm$\,0.0065 & 1.000\,$\pm$\,0.020 & 1.000\,$\pm$\,0.027 & 1.000\,$\pm$\,0.043 \\
{He\,\textsc{i}\,$\lambda$4922} & 0.0056\,$\pm$\,0.0018 & \nodata & \nodata & \nodata & \nodata \\
{[O\,\textsc{iii}]\,$\lambda$4959} & 1.4318\,$\pm$\,0.0053 & 0.8482\,$\pm$\,0.0042 & 1.504\,$\pm$\,0.021 & 0.612\,$\pm$\,0.023 & 1.070\,$\pm$\,0.045 \\
{[O\,\textsc{iii}]\,$\lambda$5007} & 4.337\,$\pm$\,0.011 & 2.5839\,$\pm$\,0.0077 & 4.796\,$\pm$\,0.032 & 1.806\,$\pm$\,0.034 & 2.393\,$\pm$\,0.077 \\
{He\,\textsc{i}\,$\lambda$5015} & 0.0291\,$\pm$\,0.0021 & 0.0240\,$\pm$\,0.0025 & \nodata & \nodata & \nodata \\
{He\,\textsc{i}\,$\lambda$5876} & \nodata & 0.0532\,$\pm$\,0.0093 & \nodata & \nodata & \nodata \\
{[N\,\textsc{ii}]\,$\lambda$6548} & \nodata & 0.0296\,$\pm$\,0.0077 & 0.0211\,$\pm$\,0.0063 & 0.0060\,$\pm$\,0.0017 & \nodata \\
{[H$\alpha$\,$\lambda$6563} & 2.786\,$\pm$\,0.047 & 2.785\,$\pm$\,0.049 & 2.860\,$\pm$\,0.058 & 2.860\,$\pm$\,0.047 & 2.86\,$\pm$\,0.15 \\
{[N\,\textsc{ii}]\,$\lambda$6584} & 0.0163\,$\pm$\,0.0047 & 0.0724\,$\pm$\,0.0086 & 0.0634\,$\pm$\,0.0010 & 0.01798\,$\pm$\,0.00029 & \nodata \\
{He\,\textsc{i}\,$\lambda$6678} & 0.0284\,$\pm$\,0.0046 & \nodata & \nodata & \nodata & \nodata \\
{[S\,\textsc{ii}]\,$\lambda$6717} & 0.0653\,$\pm$\,0.0049 & 0.137\,$\pm$\,0.010 & 0.176\,$\pm$\,0.021 & 0.0775\,$\pm$\,0.0053 & 0.076\,$\pm$\,0.098 \\
{[S\,\textsc{ii}]\,$\lambda$6731} & 0.0436\,$\pm$\,0.0063 & 0.100\,$\pm$\,0.012 & 0.090\,$\pm$\,0.027 & 0.0564\,$\pm$\,0.0077 & 0.222\,$\pm$\,0.099 \\
{He\,\textsc{i}\,$\lambda$7065} & 0.0037\,$\pm$\,0.0052 & 0.021\,$\pm$\,0.011 & \nodata & \nodata & \nodata \\
\hline
F(H$\beta$) ($\times$10$^{-17}$\,erg\,s$^{-1}$\,cm$^{-2}$) & 181.91\,$\pm$\,0.89 & 220.0\,$\pm$\,1.4 & 140.7\,$\pm$\,2.8 & 172.0\,$\pm$\,4.6 & 206.1\,$\pm$\,8.8 \\
EW(H$\beta$) (\AA) & 96.0\,$\pm$\,1.3 & 47.51\,$\pm$\,0.44 & 169\,$\pm$\,27 & 66.4\,$\pm$\,5.5 & 43.5\,$\pm$\,4.4 \\
c(H$\beta$) & 0.001 & 0.001 & 0.056 & 0.001 & 0.501 \\
EW$(a_{\textnormal{H}\textsc{i}}$) (\AA) & 4.50 & 3.44 & 4.43 & 3.52 & 2.47 \\
\enddata 
\tablecomments{Measured emission line fluxes, corrected for underlying stellar absorption and internal reddening, for some objects of our BCD sample. The equivalent width of the underlying stellar absorption is reported at H$\beta$. Emission line intensities for the full sample of BCDs are available online.}
\label{tab:flux}
\end{deluxetable*}

%The measured emission line fluxes are then corrected for foreground extinction in the line of sight of the observed object, using the Galactic dust reddening map by Schlafy and Finkbeiner \citep{2011ApJ...737..103S} and $E(B-V)$\,=\,$A_{V}$\,/\,3.1, and an SMC reddening curve. We then correct for internal reddening by scaling the observed Balmer line ratios to the theoretical Balmer line ratios for an \HII\ region of $T_{e}$\,=\,10,000~K in the high ionization zone. The flux estimates and corresponding errors, corrected for reddening, are reported in the online tables and for our sample of BCDs observed at Lick and Keck, respectively.

%\begin{itemize}
%\item correct for Galactic extinction in the direction of the observed object using Schlafly and Finkbeiner (2011)'s dust map
%\item E(B-V) from SandF
%\item assume E(B-V) = A$_{V}$/3.1
%\item calculate the A$_{\lambda}$ from A$_{V}$ at emission lines of interest
%\item measure emission line fluxes uses Absorption LIne Software (ALIS; cite)
%\item emission line flux ratios reported in Table
%\item ...
%\item correct spectra for underlying stellar absorption; Balmer decrement????!!!!
%\end{itemize}

\subsection{Metallicity}
Our sample of observed BCDs consists of six systems confirmed or predicted to have metallicities in the lowest-metallicity regime, with gas phase oxygen abundance 12+log(O/H)\,$\lesssim$\,7.20, or $Z$\,$\lesssim$\,0.03\,$Z_{\odot}$. These systems are listed in Table \ref{tab:lowestmetallicity} with their metallicities and the method by which we obtained a measurement of their gas phase oxygen abundance. We are able to obtain an empirical estimate of the metallicity using the $R$ and $S$ methods on systems observed using Shane$+$Kast, or obtain a direct measurement of the metallicity using the temperature sensitive oxygen line at [\OIII] $\lambda$4363\,\AA\ on systems observed using Keck$+$LRIS. The following sections describe these methods in more detail.

\begin{deluxetable}{ccc}
\tablewidth{0pt}%\textwidth}
\tablecaption{BCDs in the Lowest-Metallicity Regime} 
\tablehead{
\colhead{Target Name} & \colhead{12\,+\,log(O/H)} & \colhead{Metallicity Method}}
\startdata
J0018$+$2345 & 7.18\,$\pm$\,0.03 & $R$\,and\,$S$ \\
J0834$+$5905 & 7.17\,$\pm$\,0.13 & Direct \\
Little Cub & 7.13\,$\pm$\,0.08 & Direct \\
J1045$+$0209 & 6.48\,$\pm$\,0.31 & $R$\,and\,$S$ \\
J1214$+$1245 & 7.17\,$\pm$\,0.13 & $R$\,and\,$S$ \\
J1554$+$4620 & 7.24\,$\pm$\,0.09 & $R$\,and\,$S$ \\
\hline
J0943$+$3326 & 7.16\,$\pm$\,0.07 & Direct \\
\enddata
\tablecomments{The six systems in our sample that are either confirmed or predicted to have metallicities in the lowest-metallicity regime, 12\,+\,log(O/H)\,$\lesssim$\,7.20. We note that the metallicity measurement of J0834$+$5905 is a lower limit of its true metallicity. We note that we also list J0943$+$3326 here, known in the literature as AGC198691 \citep{2016ApJ...822..108H}. Our survey independently identified this system as a candidate metal-poor galaxy, and the values reported here reflect our measurements. Since this galaxy was first reported by \citet{2016ApJ...822..108H}, we do not include it as one of the six lowest-metallicity systems identified by this survey.}
\label{tab:lowestmetallicity}
\end{deluxetable}

\subsubsection{Lick Data}
The temperature sensitive oxygen line at [\OIII] $\lambda$4363\,\AA, which is necessary for a direct abundance measurement, is typically not detected in our sample of BCDs observed using the Kast spectrograph owing to the lower S/N of those spectra. We therefore rely on empirical methods to estimate the metallicity of our candidate BCDs with 3-m observations. We adopt two separate methods for determining the oxygen abundance in \HII\ regions, each using the intensities, $I$, of three strong emission lines, as presented by \citet{2016MNRAS.457.3678P}. The $R$ calibration uses the intensities of $R_{2}$, $R_{3}$, and $N_{2}$ and the $S$ calibration uses the intensities of $S_{2}$, $R_{3}$, and $N_{2}$, where the standard notations are:

\begin{eqnarray}
\begin{array}{ccc}
R_{2}\,=\,I_{[\textnormal{\OII}]}\lambda\lambda3727,3729\,/\,I_{\textnormal{H}\beta} \\
N_{2}\,=\,I_{[\textnormal{\NII}]} \lambda\lambda6548,6583\,/\,I_{\textnormal{H}\beta} \\
S_{2}\,=\,I_{[\textnormal{\SII}]}\lambda\lambda6717,6731\,/\,I_{\textnormal{H}\beta} \\
R_{3}\,=\,I_{[\textnormal{\OIII}]}\lambda\lambda4959,5007\,/\,I_{\textnormal{H}\beta}
\end{array}
\label{equation:stronglines}
\end{eqnarray}

\noindent The $R$ and $S$ calibrations are bifurcated; the oxygen abundance is estimated from either the lower or the upper branch depending on the value of log(\textit{N$_{2}$}). The lower branch is used for \HII\ regions with log(\textit{N$_{2}$})\,$<$\,--0.6:
\begin{eqnarray}
\begin{array}{ccc}
{\rm 12+log(O/H)}_{R,L}\,=\,7.932\,+\,0.944\,\log(R_{3}/R_{2})\,+\,0.695\,\log\,N_{2} \\  
+\,(0.970\,-\,0.291\,\log(R_{3}/R_{2})\,-\,0.019\,\log\,N_{2}) \\
\,\times\,\log\,R_{2} \\ 
\end{array}
\label{equation:RL}
\end{eqnarray}
\begin{eqnarray}
\begin{array}{ccc}
{\rm 12+log(O/H)}_{S,L}\,=\,8.072\,+\,0.789\,\log(R_{3}/S_{2})\,+\,0.726\,\log\,N_{2} \\  
+\,(1.069\,-\,0.170\,\log(R_{3}/S_{2})\,+\,0.022\,\log\,N_{2}) \\
\,\times\,\log\,S_{2} \\ 
\end{array}
\label{equation:SL}
\end{eqnarray}
The upper branch is  applicable  for \HII\ regions with log\textit{N$_{2}$}\,$\geq$\,--0.6:
\begin{eqnarray}
\begin{array}{ccc}
{\rm 12+log(O/H)}_{R,U}\,=\,8.589\,+\,0.022\,\log(R_{3}/R_{2})\,+\,0.399\,\log\,N_{2} \\  
+\,(-0.137\,+\,0.164\,\log(R_{3}/R_{2})\,+\,0.589\,\log\,N_{2}) \\
\,\times\,\log\,R_{2} \\ 
\end{array}
\label{equation:RU}
\end{eqnarray}
\begin{eqnarray}
\begin{array}{ccc}
{\rm 12+log(O/H)}_{S,U}\,=\,8.424\,+\,0.030\,\log(R_{3}/S_{2})\,+\,0.751\,\log\,N_{2} \\  
+\,(-0.349\,+\,0.182\,\log(R_{3}/S_{2})\,+\,0.508\,\log\,N_{2}) \\
\,\times\,\log\,S_{2} \\ 
\end{array}
\label{equation:SU}
\end{eqnarray}

%For comparison, we also compute metallicities using \textsc{PyNeb} for our sample observed using the Kast spectrograph. Computations using \textsc{PyNeb} assume a two-zone photoionization model of the \HII\ region in these BCDs and assume a high ionization zone temperature of $T_{e}$\,=\,10,000~K for all the systems, which is typical of \HII\ regions, and present 

For systems where we do not detect the weaker metal lines, [\NII] and/or [\SII], we adopt a 3$\sigma$ upper limit on their fluxes in order to estimate their metallicities. The reported metallicity of each galaxy in our BCD sample is based on the mean oxygen abundance derived from the $R$ and $S$ calibrations. We note that the separate $R$ and $S$ metallicity estimates are often in good agreement with one another, with the mean and standard deviation of the absolute value difference between the two methods, $|~$12\,+\,log(O/H)$_{R}$\,--\,12\,+\,log(O/H)$_{S}~|$, being 0.055\,$\pm$\,0.179. Resulting values are listed in Table \ref{tab:observations}, with the full sample available online.

\subsubsection{Keck Data}
The data acquired using LRIS at Keck Observatory are of much higher S/N and allow for both density- and temperature-sensitive emission lines to be detected. All calculations of the electron density, electron temperature, ionic abundances, and resulting metallicities were made using \textsc{PyNeb} \citep{2015A&A...573A..42L}.\footnote{\textsc{PyNeb} is available from: \url{http://www.iac.es/proyecto/PyNeb/}}

We significantly detect the [\SII] $\lambda\lambda$6717,6731\,\AA\,doublet in all LRIS observations and use the ratio of the two lines to calculate the electron density. However, consistent with the expected electron density of an \HII\ region, the measured electron densities occupy the low-density regime, where the ratio of the [\SII] lines is less sensitive to the true electron density. Therefore, in all calculations of the metallicity, we assume a value of $n_{e}$ = 100~cm$^{-3}$ in our ionic abundance estimates, which is consistent with both the density as determined by the [\SII] $\lambda\lambda$6717,6731\,\AA\,lines and the expected range of densities in \HII\ regions, $10^{2}\,\leq\,n_{e}~\,(\textnormal{cm}^{-3})\,\leq\,10^{4}$ \citep{1989agna.book.....O}.

We assume a two-zone photoionization model of the \HII\ region in these BCDs and calculate the corresponding temperatures of the separate high and low ionization zones. The ratio of the [\OIII] $\lambda$4363\,\AA\,line to the [\OIII] $\lambda$5007\,\AA\,line allows for a determination of the temperature of the high ionization zone ($T_{e}\,$[\OIII]). We note that the temperature-sensitive oxygen line at [\OIII] $\lambda$4363\,\AA\,is detected in most of our LRIS observations, however, we adopt a 3$\sigma$ upper limit on the measured emission line flux at [\OIII] $\lambda$4363\,\AA\,when we do not significantly detect the line. This measurement allows for an estimate of the electron temperature and therefore a direct measurement of the gas phase oxygen abundance. Because we do not detect the [\OII] $\lambda\lambda$7320,7330\,\AA\,or the [\NII] $\lambda$5755\,\AA\,lines necessary for a direct measurement of the temperature in the low ionization zone ($T_{e}\,$[\OII]), we adopt the formulation relating the two temperatures presented by \citet{1992MNRAS.255..325P}:

\begin{equation}
t_{e}^{-1}\,[\textnormal{\OII}]\,=\,0.5\,(\,t_{e}^{-1}\,[\textnormal{\OIII}]\,+\,0.8\,)
\label{equation:Te_lowion}
\end{equation}

where $t_{e}$\,=\,$T_{e}$\,/\,10$^{4}$~K. Because this relation is derived from modeling of photoionized regions, we perturb the calculated low ionization zone temperature by $\leq\pm$500~K to account for the systematic uncertainty in the conversion, where 500~K is the 1$\sigma$ uncertainty from the spread in the models.

The two-zone photoionization model of the \HII\ region also assumes that the total oxygen abundance is the sum of the singly and doubly ionized states:
\begin{equation} 
\frac{\textnormal{O}}{\textnormal{H}}=\,\frac{\textnormal{O}^{+}}{\textnormal{H}^{+}}\,+\,\frac{\textnormal{O}^{++}}{\textnormal{H}^{+}}
\end{equation}
The measurements of electron density, electron temperature, ionic abundances, and oxygen abundances of our Keck BCD sample are presented in Table \ref{tab:keckmetallicity}. This Keck BCD sample will appear in full, i.e., their spectra and further analysis, in a forthcoming work.

\begin{deluxetable*}{ccccccc}
\tablewidth{0pt}
\tablecaption{Physical and Chemical Properties of the BCDs Observed with Keck} 
\tablehead{ 
\colhead{Target Name} & \colhead{$n_{e}$([\SII])} & \colhead{$T_{e}$([\OIII])} & \colhead{$T_{e}$([\OII])} & \colhead{O$^{++}$\,/\,H$^{+}$} & \colhead{O$^{+}$\,/\,H$^{+}$} & \colhead{12\,+\,log(O/H)} \\
\colhead{} & \colhead{(cm$^{-3}$)} & \colhead{(K)} & \colhead{(K)} & \colhead{($\times$10$^{-6}$)} & \colhead{($\times$10$^{-6}$)} & \colhead{}
}
\startdata 
J0000+3052A & 120\,$^{+115}_{-65}$ & 15130\,$\pm$\,190 & 13671\,$\pm$\,78 & 45.2\,$\pm$\,1.4 & 7.01\,$\pm$\,0.17 & 7.72\,$\pm$\,0.01 \\ 
J0000+3052B & 150\,$^{+140}_{-76}$ & 15190\,$\pm$\,350 & 13670\,$\pm$\,140 & 26.7\,$\pm$\,1.5 & 15.86\,$\pm$\,0.58 & 7.63\,$\pm$\,0.02 \\ 
J0118+3512 & 116\,$^{+46}_{-42}$ & 15500\,$\pm$\,190 & 14000\,$\pm$\,76 & 29.27\,$\pm$\,0.84 & 8.50\,$\pm$\,0.18 & 7.58\,$\pm$\,0.01 \\ 
J0140+2951 & 18\,$^{+16}_{-11}$ & 12200\,$\pm$\,29 & 11856\,$\pm$\,15 & 88.39\,$\pm$\,0.71 & 22.97\,$\pm$\,0.22 & 8.05\,$\pm$\,0.00 \\ 
J0201+0919 & 74\,$^{+70}_{-40}$ & 14730\,$\pm$\,440 & 13850\,$\pm$\,190 & 39.8\,$\pm$\,3.1 & 11.29\,$\pm$\,0.53 & 7.71\,$\pm$\,0.03 \\ 
J0220+2044A & 135\,$^{+91}_{-71}$ & 15900\,$\pm$\,440 & 13540\,$\pm$\,170 & 25.2\,$\pm$\,1.6 & 8.65\,$\pm$\,0.38 & 7.53\,$\pm$\,0.03 \\ 
J0220+2044B & \nodata & 17500\,$\pm$\,1100 & 15060\,$\pm$\,380 & 20.5\,$\pm$\,3.0 & 3.98\,$\pm$\,0.37 & 7.39\,$\pm$\,0.06 \\ 
J0452--0541 & 42\,$^{+39}_{-28}$ & 15490\,$\pm$\,410 & 13570\,$\pm$\,160 & 27.2\,$\pm$\,1.7 & 16.01\,$\pm$\,0.66 & 7.64\,$\pm$\,0.02 \\ 
J0743+4807 & 81\,$^{+80}_{-55}$ & 9500\,$\pm$\,1600 & 10200\,$\pm$\,1100 & $\geq$\,250 & $\geq$\,90 & $\geq$\,8.34 \\ 
J0812+4836 & 70\,$^{+71}_{-35}$ & 17200\,$\pm$\,4000 & 14500\,$\pm$\,1600 & $\geq$\,11 & $\geq$\,17 & $\geq$\,7.35 \\ 
J0834+5905 & 350\,$^{+390}_{-190}$ & 21000\,$\pm$\,3100 & 15480\,$\pm$\,980 & $\geq$\,7.7 & $\geq$\,7.8 & $\geq$\,7.17 \\ 
KJ5 & 170\,$^{+118}_{-90}$ & 11570\,$\pm$\,560 & 11670\,$\pm$\,300 & 68.0\,$\pm$\,11.0 & 12.5\,$\pm$\,1.3 & 7.90\,$\pm$\,0.06 \\ 
KJ5B & 125\,$^{+100}_{-68}$ & 14030\,$\pm$\,390 & 13310\,$\pm$\,180 & 40.7\,$\pm$\,3.1 & 10.57\,$\pm$\,0.51 & 7.71\,$\pm$\,0.03 \\ 
J0943+3326 & 330\,$^{+320}_{-150}$ & 16500\,$\pm$\,1300 & 14700\,$\pm$\,500 & 10.4\,$\pm$\,2.1 & 4.06\,$\pm$\,0.47 & 7.16\,$\pm$\,0.07 \\ 
Little Cub & 32\,$^{+34}_{-17}$ & 18600\,$\pm$\,2200 & 14680\,$\pm$\,720 & 5.1\,$\pm$\,1.5 & 9.1\,$\pm$\,1.6 & 7.13\,$\pm$\,0.08 \\ 
KJ97 & 48\,$^{+44}_{-24}$ & 11880\,$\pm$\,310 & 12450\,$\pm$\,160 & 72.2\,$\pm$\,5.9 & 28.7\,$\pm$\,1.4 & 8.00\,$\pm$\,0.03 \\ 
KJ29 & 900\,$^{+640}_{-390}$ & 14270\,$\pm$\,340 & 13370\,$\pm$\,150 & 29.2\,$\pm$\,1.9 & 12.72\,$\pm$\,0.49 & 7.62\,$\pm$\,0.02 \\ 
KJ2 & 450\,$^{+450}_{-240}$ & 17550\,$\pm$\,150 & 14907\,$\pm$\,52 & 24.40\,$\pm$\,0.48 & 2.466\,$\pm$\,0.042 & 7.43\,$\pm$\,0.01 \\ 
J1414--0208 & 112\,$^{+100}_{-66}$ & 14700\,$\pm$\,1400 & 13520\,$\pm$\,610 & 19.1\,$\pm$\,5.6 & 13.5\,$\pm$\,2.3 & 7.50\,$\pm$\,0.09 \\ 
J1425+4441 & 180\,$^{+180}_{-100}$ & 15070\,$\pm$\,960 & 13320\,$\pm$\,400 & 18.2\,$\pm$\,3.0 & 13.5\,$\pm$\,1.4 & 7.50\,$\pm$\,0.06 \\ 
J1655+6337 & \nodata & 16620\,$\pm$\,160 & 13914\,$\pm$\,59 & 21.90\,$\pm$\,0.49 & 5.169\,$\pm$\,0.093 & 7.43\,$\pm$\,0.01 \\ 
J1705+3527 & 83\,$^{+52}_{-44}$& 15510\,$\pm$\,130 & 13453\,$\pm$\,54 & 37.69\,$\pm$\,0.77 & 7.69\,$\pm$\,0.14 & 7.66\,$\pm$\,0.01 \\ 
J1732+4452 & 297\,$^{+92}_{-78}$ & 15200\,$\pm$\,240 & 14146\,$\pm$\,98 & 32.9\,$\pm$\,1.3 & 8.94\,$\pm$\,0.23 & 7.62\,$\pm$\,0.02 \\ 
J1757+6454 & 69\,$^{+41}_{-34}$ & 14480\,$\pm$\,190 & 13451\,$\pm$\,81 & 39.1\,$\pm$\,1.4 & 13.67\,$\pm$\,0.34 & 7.72\,$\pm$\,0.01 \\ 
J2030--1343 & 25\,$^{+21}_{-15}$ & 13890\,$\pm$\,140 & 13446\,$\pm$\,64 & 55.9\,$\pm$\,1.6 & 13.54\,$\pm$\,0.26 & 7.84\,$\pm$\,0.01 \\ 
J2213+1722 & 29\,$^{+25}_{-17}$ & 15420\,$\pm$\,140 & 13400\,$\pm$\,58 & 30.18\,$\pm$\,0.67 & 10.20\,$\pm$\,0.18 & 7.61\,$\pm$\,0.01 \\ 
J2230--0531 & 77\,$^{+40}_{-36}$ & 14860\,$\pm$\,170 & 13845\,$\pm$\,72 & 32.49\,$\pm$\,1.00 & 8.79\,$\pm$\,0.18 & 7.62\,$\pm$\,0.01 \\ 
J2319+1616 & 137\,$^{+39}_{-39}$ & 10617\,$\pm$\,23 & 11628\,$\pm$\,13 & 106.82\,$\pm$\,0.80 & 44.79\,$\pm$\,0.42 & 8.18\,$\pm$\,0.00 \\ 
J2339+3230 & 15.3\,$^{+20.7}_{-9.1}$ & 13990\,$\pm$\,240 & 13470\,$\pm$\,110 & 50.7\,$\pm$\,2.3 & 12.17\,$\pm$\,0.36 & 7.80\,$\pm$\,0.02 \\
\enddata
\label{tab:keckmetallicity}
\tablecomments{Measurements of the electron density, electron temperature, ionic abundances, and element abundances of our sample observed with Keck+LRIS. All calculations are made using \textsc{PyNeb}. Calculations of the electron temperature and abundances assume an electron density of $n_{e}$\,=\,100\,cm$^{-3}$ due to the density insensitivity of the [\SII] $\lambda$6716/$\lambda$6731 line in the low density regime. All systems have direct metallicity estimates, except for J0743$+$4807, J0812$+$4836, and J0834$+$5905, where we do not significantly detect the [\OIII] $\lambda$4363\,\AA\,line and adopt an upper limit to the [\OIII] $\lambda$4363\,\AA\,emission line flux equivalent to three times the error of the measured line at that wavelength. In these cases, the resulting ionic abundances and metallicities are lower limits. The objects prefixed with KJ were also observed by \citet{2017MNRAS.465.3977J}. J0943$+$3326 is also known in the literature as AGC198691 \citep{2016ApJ...822..108H}. Our survey independently identified this system as a candidate metal-poor galaxy, and the values reported here reflect our measurements.}
\end{deluxetable*}

\subsubsection{$R$ and $S$ Calibration versus Direct Metallicity Measurements}
\begin{figure}
\centering
% made in ~/BCDs/telescope_proposal/2018A_Keck_Bolte/figure_making.ipynb
\includegraphics[width=0.5\textwidth]{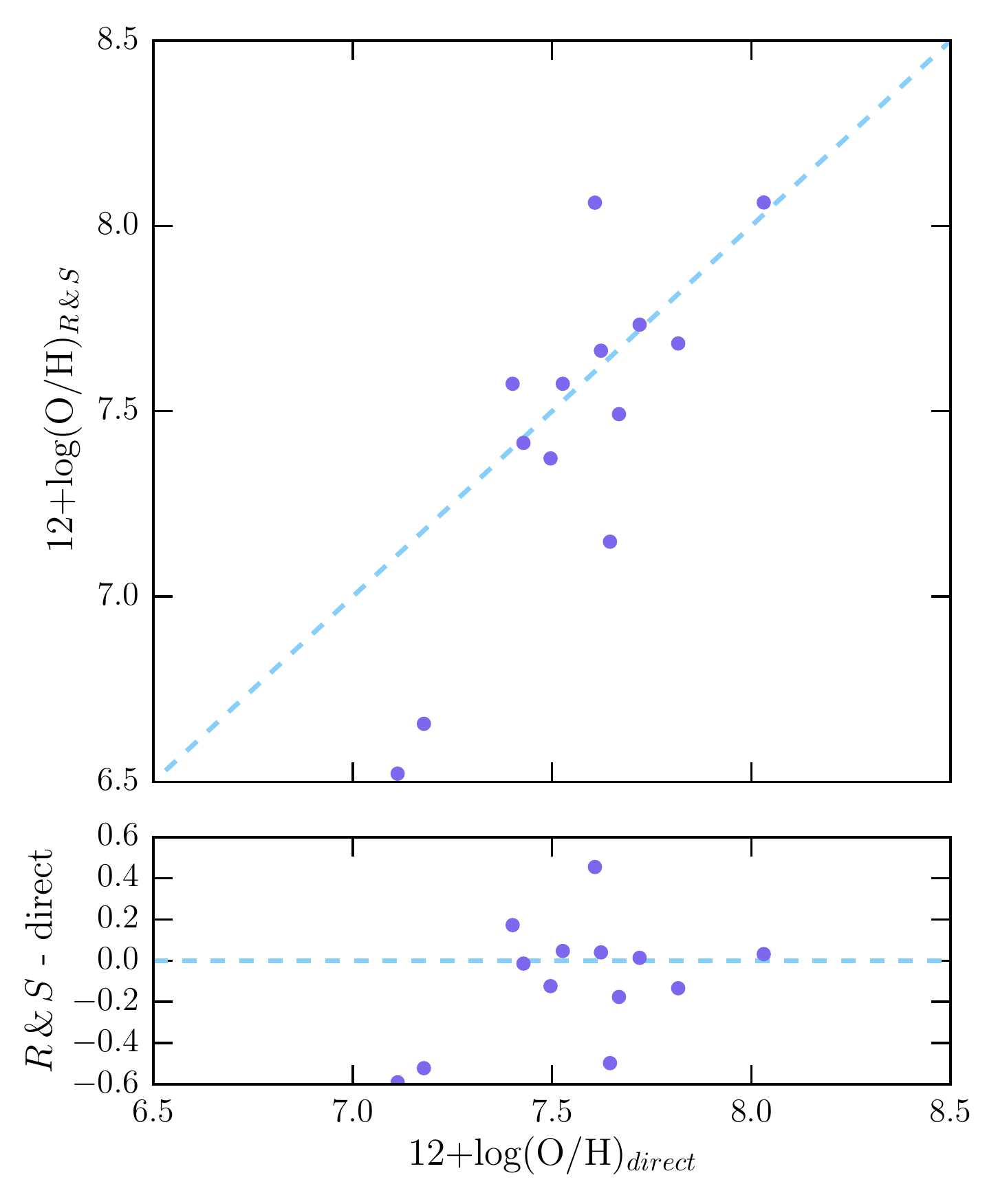}
\caption{Comparison between the $R$ and $S$ metallicity estimates and direct metallicity measurements for thirteen BCDs for which we obtained both Kast and LRIS spectra. The upper panel shows the direct versus $R$ and $S$ calibration metallicities for each system (purple points) and the one-to-one relation between the two measurements (dashed blue line). The lower panel shows how much the $R$ and $S$ calibration methods over- or under-estimated the true metallicity.}
\label{fig:OHcomparison}
\end{figure}

Our sample contains thirteen BCDs for which we obtained both Kast and LRIS spectra. Using these systems, we consider the reliability of the $R$ and $S$ calibration methods in providing a reasonable estimate of the metallicity of the system measured via the direct method. In the upper panel of Figure \ref{fig:OHcomparison}, we show the direct metallicity measurements versus $R$ and $S$ calibration estimates of the metallicity for the thirteen systems, along with the idealized one-to-one scenario where the calibration method exactly predicts the direct metallicity. We calculate the 12\,+\,log(O/H)$_{direct}$\,--\,12\,+\,log(O/H)$_{R\&S}$ of these thirteen BCDs, shown in the lower panel. The mean and standard deviation of the difference between the two metallicities is 0.010$\,\pm\,$0.284 dex.

The $R$ and $S$ calibration methods presented by \citet{2016MNRAS.457.3678P} were derived using a compilation of 313 \HII\ regions with direct metallicity measurements. Their sample has a mean oxygen abundance of 12\,+\,log(O/H)$\,\sim\,$8.0 and only a small fraction of their sample occupied the low metallicity regime at 12\,+\,log(O/H)$\,\leq\,$7.65, which may cause the resulting relations to be less well-calibrated at the low metallicity regime. In our sample of thirteen BCDs, the two systems that occupy the lowest metallicity regime at 12\,+\,log(O/H)$_{direct}\,\lesssim\,$7.20 had metallicities significantly underestimated using the $R$ and $S$ calibration, 12\,+\,log(O/H)$_{R\&S}\,\sim\,$6.60. While it is possible that some of the systems in our sample with 12\,+\,log(O/H)$_{R\&S}\,\leq\,$7.0 have underestimated metallicities, there is a monotonic trend in that the systems predicted to be of the lowest metallicities using the $R$ and $S$ calibrations remain as the lowest metallicity systems of our sample. This bolsters our confidence in being able to identify the lowest metallicity systems from the strong line $R$ and $S$ calibration methods for follow-up observations and direct metallicity measurements.

\subsection{Derived Properties: Distance, H$\alpha$ Luminosity, and Star Formation Rate}
\label{distance+props}
% made in ~/BCDs/papers/kast_survey/make_plots_tables
\begin{figure}
\centering
\includegraphics[width=0.5\textwidth]{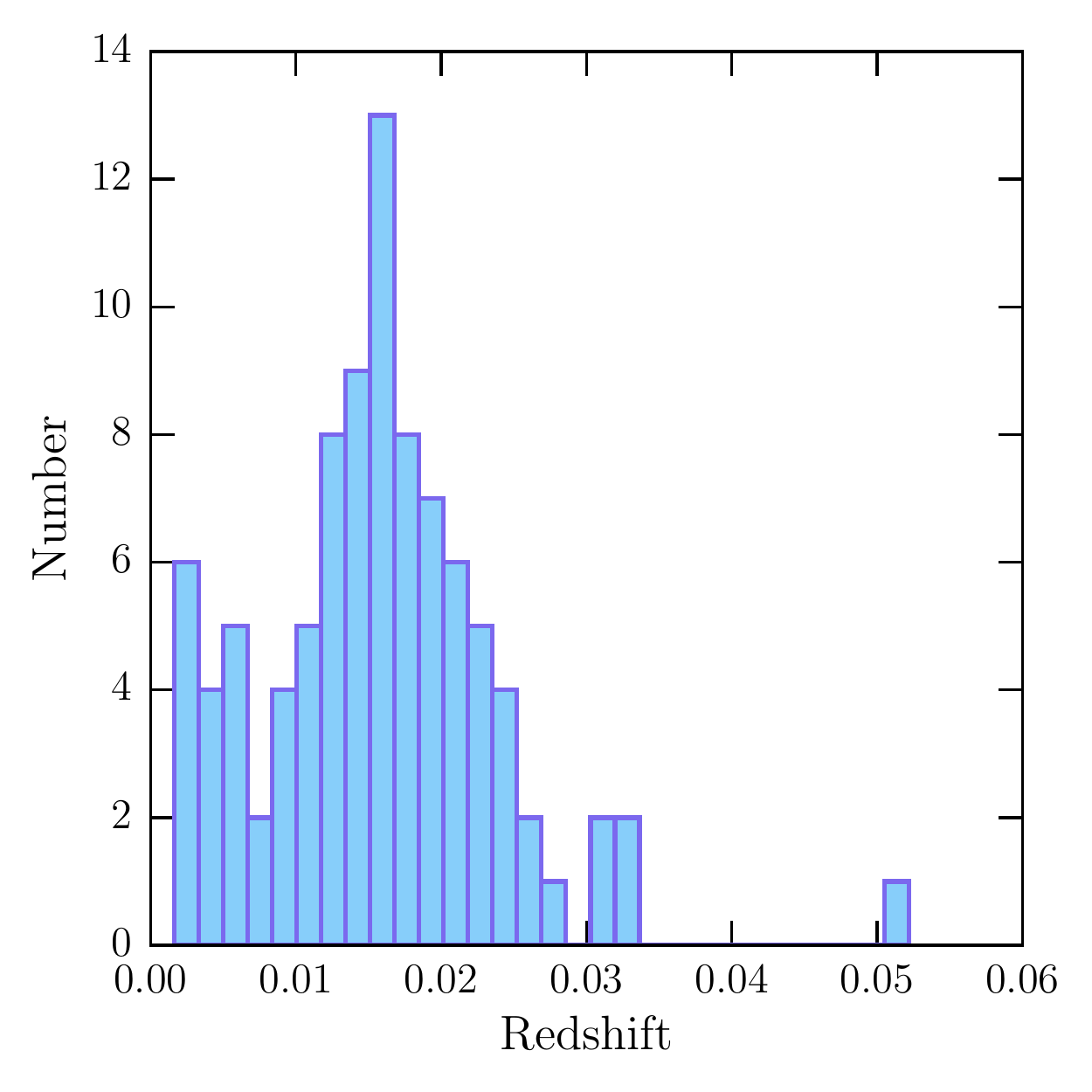}
\caption{The redshift distribution of our full sample of BCDs. The mean redshift of our sample is $z$\,=\,0.016, which corresponds to a luminosity distance of 70.6 Mpc in a \textit{Planck} cosmology. Our highest redshift object has $z$\,=\,0.052.}
\label{fig:redshift}
\end{figure}

We show a redshift distribution of our full sample of BCDs in Figure \ref{fig:redshift}. Using these measured redshifts, we calculate the luminosity distance ($d_{L}$) to each system using \textsc{astropy}'s cosmology subpackage, assuming the built in \textsc{Planck15} cosmology \citep{2016AA...594A..13P}. The values are reported for a subsample in Table \ref{tab:derivedprops} and available in its entirety online. However, these distance measurements are not well constrained with our available data given the local velocity field. For comparison, we include an additional estimate of the distance using the \citet{2000ApJ...529..786M} flow model, which corrects for the local velocity field. We note that flow model estimates can be highly uncertain for nearby galaxies, and more reliable distance measurements would require additional data, such as photometry of the tip of the red giant branch (TRGB). 

%%%ha_lum = kl['H_alpha']*1e-17*u.erg/u.s/(u.cm**2.) * \
%%%            (4.*np.pi*((kl['luminosity distance']).to(u.cm)**2.))

Nevertheless, for completeness, we adopt the luminosity distances and calculate distance-dependent properties for each system and list these values in Table \ref{tab:derivedprops}, with the caveat that these quantities depend on the somewhat uncertain distance estimates. The reported H$\alpha$ luminosity of each system, $L$(H$\alpha$), is calculated using our observed H$\alpha$ fluxes combined with the assumed distance determined above:
\begin{equation} 
\textnormal{$L$(H$\alpha$)}\,=\,\textnormal{$F$(H$\alpha$)}~4\pi d_{L}^{2}
\end{equation}
%where the H$\alpha$ flux is in units of 10$^{-17}$\,erg\,s$^{-1}$\,cm$^{-2}$ and $d_{L}$ is the luminosity distance, in centimeters.
The resulting star formation rate (SFR) is calculated using the \citeauthor{1998ARA&A..36..189K} relation between $L$(H$\alpha$) and SFR:
\begin{equation} 
\textnormal{SFR}\,=\,7.9\,\times\, 10^{-42}\,L(\textnormal{H}\alpha)
\end{equation}
where the SFR is in units of $M_{\odot}\,\textnormal{year}^{-1}$ and $L$(H$\alpha$) in erg\,s$^{-1}$. We then divide this SFR by a factor of 1.8, which corrects for the flattening of the stellar initial mass function (IMF) below 1\,$M_{\odot}$ for a \citet{2003PASP..115..763C} IMF, instead of the power law Salpeter IMF adopted by \citet{1998ARA&A..36..189K}.

We note that the \citet{1998ARA&A..36..189K} calibration between $L$(H$\alpha$) and the SFR is based on measurements of more metal-rich systems than the BCDs considered in this sample, which adds uncertainty in the calculation of a SFR from $L$(H$\alpha$). In particular, massive O and B stars in low metallicity environments are likely more efficient at ionizing their surroundings than their metal-rich counterparts, meaning that the presented SFR may be an overestimate of the true SFR of the galaxy. It is also possible that in some of our BCDs, the IMF is not well-sampled, which would also lead to a deviation from the \citet{1998ARA&A..36..189K} relation.

\subsection{Stellar Mass}
\label{stellar_mass}
We present estimates of the stellar mass of each BCD using the stellar mass-to-light ($M/L$) ratios presented in \citet{2003ApJS..149..289B}. We adopt the calibrations using the $r$- and $i$-band magnitudes, specifically the $i$-band coefficients and $r-i$ color, given below. The observed photometry of these BCDs is likely to be influenced by the strong emission lines from the \HII\ region, in addition to the light of the young O and B stars. We therefore select the bands that are least likely to be contaminated by the star-forming event.
\begin{equation} 
\textnormal{log}_{10}\Big(\frac{M}{L}\Big)\,=\,0.006\,+\,\big(1.114\,\times\,(r-i)\big)
\end{equation}
The resulting stellar mass estimates are given in short in Table \ref{tab:derivedprops} and in full online.

\begin{deluxetable*}{ccccccc}
\tablewidth{0pt}
\tablecaption{Derived Properties of our BCD sample} 
\tablehead{ 
\colhead{Target Name} & \colhead{$d_{\textnormal{L}}$}& \colhead{$d_{\textnormal{Mould}}$} & \colhead{$M_{\textit{B}}$} & \colhead{$L$(H$\alpha$)} & \colhead{SFR} & \colhead{$M_{*}$}  \\
\colhead{} & \colhead{(Mpc)} & \colhead{(Mpc)} & \colhead{} & \colhead{($\times$10$^{39}$\,erg\,s$^{-1}$)} & \colhead{($\times$10$^{-3}$\,$M_{\odot}\,\textnormal{year}^{-1}$)} & \colhead{($\times$10$^{6}$\,$M_{\odot}$)}
}
\startdata 
J0000+3052A & 67.6 & 67.1 & -14.29 & 2.77\,$\pm$\,0.16 & 12.15\,$\pm$\,0.70 & 6.6\,$\pm$\,1.1 \\
J0000+3052B & 68.5 & 68.1 & -14.52 & 3.44\,$\pm$\,0.20 & 15.11\,$\pm$\,0.87 & 17.9\,$\pm$\,3.1 \\
J0003+3339 & 94.7 & 93.6 & -15.19 & 4.31\,$\pm$\,0.20 & 18.94\,$\pm$\,0.86 & 25.8\,$\pm$\,6.6 \\
J0018+2345 & 68.8 & 67.9 & -14.83 & 2.79\,$\pm$\,0.14 & 12.22\,$\pm$\,0.63 & 21.1\,$\pm$\,3.2 \\
J0033--0934 & 54.2 & 53.4 & -15.55 & 2.07\,$\pm$\,0.12 & 9.10\,$\pm$\,0.54 & 61.0\,$\pm$\,5.6 \\
J0035--0448 & 75.9 & 74.4 & -14.71 & 2.824\,$\pm$\,0.086 & 12.39\,$\pm$\,0.38 & 19.3\,$\pm$\,3.3 \\
J0039+0120 & 66.0 & 64.7 & -14.10 & 0.813\,$\pm$\,0.033 & 3.57\,$\pm$\,0.14 & 10.5\,$\pm$\,2.2 \\
J0048+3159 & 68.5 & 67.6 & -14.96 & 0.611\,$\pm$\,0.034 & 2.68\,$\pm$\,0.15 & 35.7\,$\pm$\,8.3 \\
J0105+1243 & 63.4 & 61.9 & -14.14 & 0.681\,$\pm$\,0.035 & 2.99\,$\pm$\,0.16 & 8.6\,$\pm$\,3.3 \\
J0118+3512 & 73.9 & 72.9 & -15.00 & 6.27\,$\pm$\,0.32 & 27.5\,$\pm$\,1.4 & 25.2\,$\pm$\,3.8 \\
\enddata 
\tablecomments{We report luminosity distances for and distances corrected for the local velocity field using the \citet{2000ApJ...529..786M} flow model. Absolute $B$-band magnitudes are calculated from the empirical $ugri-UBVR_{c}$ transformations presented in \citet{2014MNRAS.445..890C}. Calculations of the H$\alpha$ luminosities, star formation rates, and stellar masses are discussed in Section \ref{AD}.}
\label{tab:derivedprops}
\end{deluxetable*}

\section{Blue Compact Dwarfs and Other Metal-Poor Systems}
\label{BCDsXMPs}

\subsection{Luminosity-Metallicity Relation}
\label{LZ_relation}
\begin{figure*}
\centering
\includegraphics[width=1.0\textwidth]{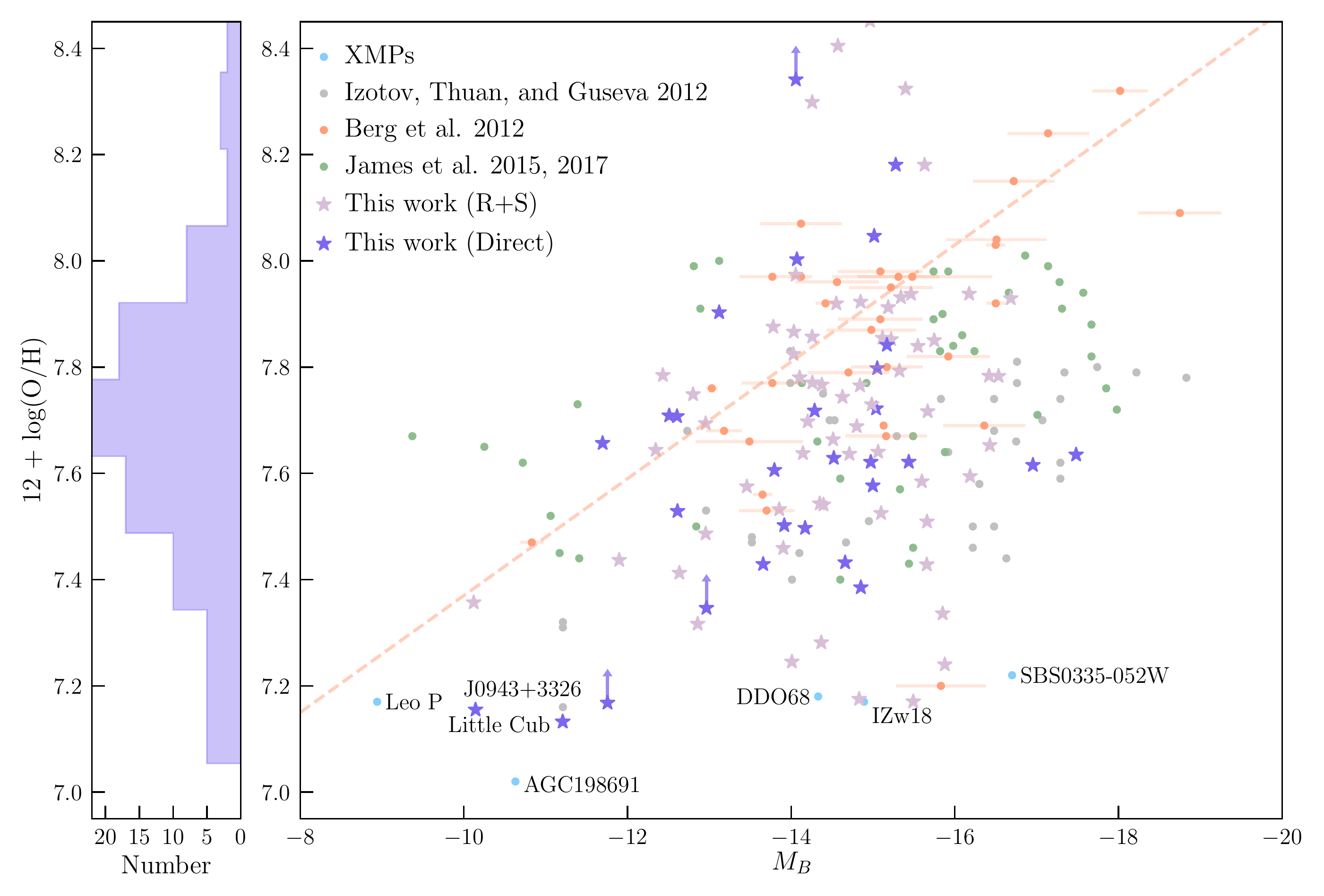}
\caption{The absolute $B$-band magnitude versus the gas phase oxygen abundance of our sample of observed BCDs, shown with star symbols, compared to several literature BCD samples. The dark purple symbols correspond to our BCDs that have a direct oxygen abundance measurement, while the light purple symbols represent BCDs with an oxygen abundance estimated via the $R$ and $S$ calibration methods. SDSS DR7 BCDs from \citet{2012A&A...546A.122I} are shown as grey points, low-luminosity star-forming galaxies from \citet{2012ApJ...754...98B} are shown in orange, and blue diffuse dwarfs from \citet{2015MNRAS.448.2687J, 2017MNRAS.465.3977J} are shown in green. Other well-known systems of extremely low metallicity are shown in blue and labeled. We note that the points labeled J0943$+$3326 and AGC198691 are the same system, with the former being measurements from our observations and the latter from the work of \citet{2016ApJ...822..108H}. The dashed orange line indicates the best fit relationship between $M_{B}$ and 12\,+\,log(O/H) as determined by \citet{2012ApJ...754...98B} and given in Equation \ref{equation:LZ}. We show the distribution of metallicities of our BCD sample in the left panel.}
\label{fig:LZ}
\end{figure*}

The luminosity-metallicity ($L-Z$) relation is thought to be a consequence of the more fundamental relation between a galaxy's mass and its chemical abundance, known as the mass-metallicity ($M-Z$) relation. %, and discussed in Section \ref{MZ_relation}.
At the low mass and low luminosity end of the relation, galaxies are more inefficient in chemically enhancing their gas and in retaining heavy metals \citep{2009A&A...505...63G}. \citet{2012ApJ...754...98B} presented a study of low luminosity galaxies with accurate distances made via the TRGB method or Cepheid observations and direct abundance measurements with the [\OIII]\,$\lambda$\,4363\,\AA\,line. Their sample showed a small scatter in the relationship between the observed luminosity and oxygen abundance, shown as the orange dashed line in Figure \ref{fig:LZ} and given by:
\begin{equation} 
12\,+\,\textnormal{log(O/H)}\,=(6.27\,\pm\,0.21)\,+\,(-0.11\,\pm\,0.01)\,M_{B}
\label{equation:LZ}
\end{equation}
Here, $M_{B}$ is the $B$-band luminosity. This relationship from the \citet{2012ApJ...754...98B} sample has a dispersion of $\sigma$\,=\,0.15.

It has been suggested that significant deviations from the $L-Z$ relation may be due to abnormal processes in the chemical evolutionary history of the galaxy and may indicate recent infall processes or disruptions that led to the observed low metallicity. \citet{2010MNRAS.406.1238E} noted that outliers of the $L-Z$ relation with \HI\ observations tend to have disrupted morphologies, suggesting that these galaxies have undergone recent or current interactions. The observed metal-poor nature of these systems is credited to the mixing of previously enhanced, more metal-rich gas with newly accreted, nearly pristine gas. Tidal interactions mix the gas and these systems are thus observed to lie below the $L-Z$ relation, i.e., have a lower metallicity than predicted by the relation, given their luminosity.

Using the empirical $ugri-UBVR_{c}$ transformations presented in \citet{2014MNRAS.445..890C}, we convert the observed SDSS magnitudes of our BCDs into absolute $B$-band luminosities:
\begin{equation} 
B-i\,=(1.27\,\pm\,0.03)\,(g-i)\,+\,(0.16\,\pm\,0.01)
\label{equation:SDSStoBband}
\end{equation}
We plot the resulting absolute $B$-band luminosity of our BCDs versus the oxygen abundance ($L-Z$ relation) in Figure \ref{fig:LZ}, and compare our results with the \citet{2012ApJ...754...98B} sample of nearby dwarf galaxies and a selection of other known low metallicity galaxies. We note that the mean residual of our sample of BCDs from the \citet{2012ApJ...754...98B}  $L-Z$ relation given in Equation \ref{equation:LZ} is 0.271; however, this is weighted by a bias towards systems that lie below the $L-Z$ relation.

A significant fraction of our BCD sample appears to be outliers of the $L-Z$ relation derived by \citet{2012ApJ...754...98B}. If the $L-Z$ relation from \citet{2012ApJ...754...98B} is representative of regular star-forming regions, i.e., chemical enrichment is a result of star-formation and subsequent feedback and enrichment from the stellar population, and deviations from this relation indicate interactions with the surrounding media, such as the inflow and accretion of pristine gas from the IGM, then it seems that there exists a larger fraction of BCDs in our sample that are experiencing recent star-formation and observed to have a low metallicity due to the accretion of metal-poor gas. This is in contrast with systems that are low metallicity simply because they have processed little of their reservoir of gas into stars since the formation of the galaxy, due to inefficient star formation. %\correct{Is this an observational bias though? Are we seeing more systems that deviate from the L-Z relation because the ones that are experiencing infall tend to be of higher luminosities? Note, though, that even our lowest direct metallicity systems are not the dimmest in apparent magnitude.}

We note that even with our sample of BCDs that have direct abundances, our distance measurements contribute a large source of uncertainty in $M_{B}$, as discussed in Section \ref{distance+props}. For BCDs with metallicities based on the $R$ and $S$ calibration methods, we must also consider the accuracy of these methods in predicting the true metallicity of a system. Therefore, in addition to the distance uncertainties, there also exists an uncertainty in the metallicity for systems that currently only afford metallicity estimates made via strong emission lines. 

Furthermore, the $B$-band flux is dominated by the light of massive O and B stars, likely on the specific population of O and B stars present. This makes the observed $B$-band luminosity more sensitive to the recent or on-going star formation and less sensitive to the stellar mass and integrated star formation history of the galaxy \citep{2005ApJ...624..661S}. The sensitivity of the $B$-band luminosity to the star formation event could shift the observed luminosity of a system to a higher luminosity than what is expected given its metallicity. Additionally, the $B$-band is also more susceptible to absorption effects than longer wavelength bands.

To make more definite conclusions about our systems and how well they follow or deviate from the \citet{2012ApJ...754...98B} $L-Z$ relation, we would require direct abundance measurements and accurate distance measurements. Alternatively, supplementary infrared imaging, which is a better proxy of galaxy mass than the $B$-band, on the sample of metal-poor BCDs could provide a more fundamental $L-Z$ analysis.

\subsection{Mass-Metallicity Relation}
\label{MZ_relation}
The stellar mass (M$_{*}$) and the metallicity of a galaxy are considered to be fundamental physical properties of galaxies and are correlated such that more massive galaxies are observed to have higher metallicities. This correlation is given by the mass-metallicity ($M-Z$) relation \citep{2010MNRAS.408.2115M, 2012ApJ...754...98B, 2015MNRAS.451.2251I, 2018AJ....155...82H}. It is unclear whether the $M-Z$ relation arises because more massive galaxies form fractionally more stars than their low-mass counterparts leading to higher metal yields \citep{2007MNRAS.375..673K}, or whether galaxies of all masses form similar fractions of stars from their gas, but low-mass galaxies subsequently lose a larger fraction of metal-enriched gas due to their shallower galactic potentials \citep{1974MNRAS.169..229L, 2004ApJ...613..898T}.

While there exists evidence for various origins of the $M-Z$ relation, both the stellar mass and metallicity track the evolution of galaxies; the stellar mass indicates the amount of gas in a galaxy trapped in the form of stars, and the metallicity of a galaxy indicates the reprocessing of gas by stars as well as any transfer of gas from the galaxy to its surrounding environment \citep{2004ApJ...613..898T}. Understanding the origin of the $M-Z$ relation would provide insight into the timing and efficiency of how galaxies process their gas into stars, which is relevant in models of the chemical evolution of galaxies over all ranges of galaxy mass and redshift.

\begin{figure*}
\centering
\includegraphics[width=1.0\textwidth]{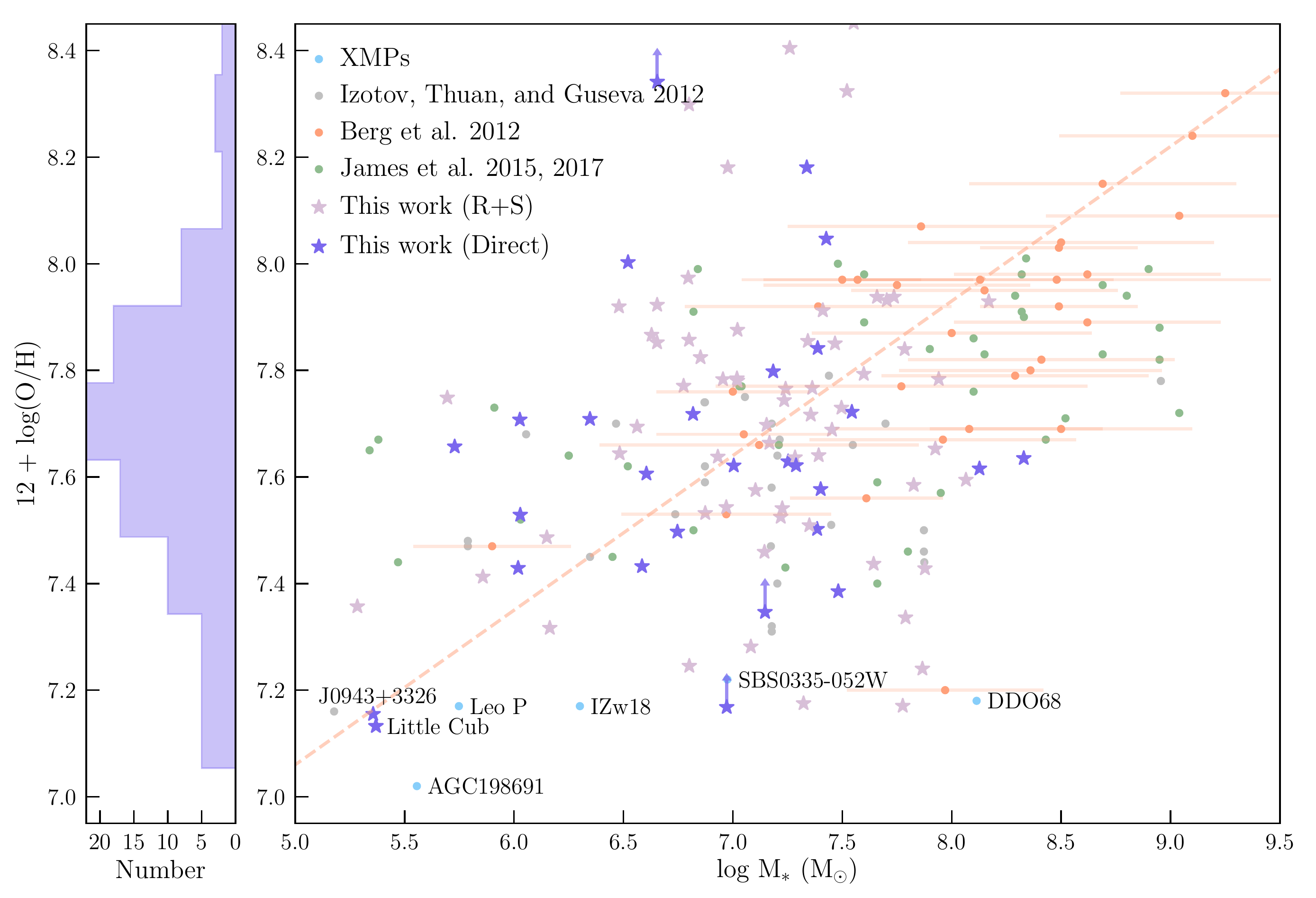}
\caption{The stellar mass versus the gas phase oxygen abundance of our sample of observed BCDs, shown with star symbols, compared to several literature BCD samples. The dark purple symbols correspond to our BCDs that have a direct oxygen abundance measurement, while the light purple symbols represent BCDs with an oxygen abundance estimated via the $R$ and $S$ calibration methods. The remaining points belong to the samples as described in Figure \ref{fig:LZ}. The dashed orange line indicates the best fit relationship between $M_{*}$ and 12\,+\,log(O/H) as determined by \citet{2012ApJ...754...98B} and given in Equation \ref{equation:MZ}. We show the distribution of metallicities of our BCD sample in the left panel.}
%SDSS DR7 BCDs from \citet{2012A&A...546A.122I} are shown as grey points, low-luminosity star-forming galaxies from \citet{2012ApJ...754...98B} are shown in orange, and blue diffuse dwarfs from \citet{2015MNRAS.448.2687J, 2017MNRAS.465.3977J} are shown in green. Other well-known systems of extremely low metallicity are shown in blue and labeled. We note that the points labeled J0943$+$3326 and AGC198691 are the same system, with the former being measurements from our observations and the latter from the work of \citet{2016ApJ...822..108H}. 
\label{fig:MZ}
\end{figure*}

Obtaining the stellar mass of a galaxy is challenging, and as a result, the luminosity of a galaxy is often adopted as a proxy of its mass. This relation is analyzed in the form of the $L-Z$ relation, as discussed previously in Section \ref{LZ_relation}. In this Section, we analyze the $M-Z$ relation in the context of our BCDs, using stellar mass estimates of our BCD sample described in Section \ref{stellar_mass}. We compare our BCDs to the \citet{2012ApJ...754...98B} $M-Z$ relation, which is:

\begin{equation} 
12\,+\,\textnormal{log(O/H)}\,=(5.61\,\pm\,0.24)\,+\,(0.29\,\pm\,0.03)\,\textnormal{log}\,(M_*)
\label{equation:MZ}
\end{equation}
We note that \citet{2012ApJ...754...98B} estimate stellar masses for their sample of low-luminosity galaxies using a combination of optical and infrared luminosities and colors: the 4.5$\mu$m luminosity, $K$\,--\,[4.5] color, and $B$\,--\,$K$ color. We direct readers to Section 6.4 of \citet{2012ApJ...754...98B} for further details. Their resulting relation has a dispersion of $\sigma$\,=\,0.15, comparable to the dispersion in their $L-Z$ relation. Our BCDs in stellar mass versus gas phase oxygen abundance space ($M-Z$ relation) are presented in Figure \ref{fig:MZ}, along with a selection of other known low metallicity galaxies.
%whereas in other studies, the $M-Z$ relation usually has a tighter dispersion.

%Studies have both found that higher mass galaxies are more gas poor, suggesting the former explanation for the $M-Z$ relation, and at the same time, galactic winds have ubiquitously been detected in starburst galaxies (Heckman, T. M. 2002, in ASP Conf. Ser. 254, Extragalactic Gas at Low Redshift, ed. J. S. Mulchaey & J. Stocke (San Francisco: ASP), 292), and absorption-line studies have consistently found metals in the intergalactic medium (Ellison, S. L., Songaila, A., Schaye, J., & Pettini, M. 2000, AJ, 120, 1175)

In addition to the uncertainty in metallicity estimates made via the $R$ and $S$ calibration methods, we must also consider that even with BCDs that afford a direct metallicity measurement, we are only able to determine the metallicity of the \HII\ region ionized by the current star formation event. Due to the massive young stars, these \HII\ regions may be self-enriched \citep{1986ApJ...300..496K}. More generally, \HII\ regions are a poor representation of BCDs as a whole since the bulk of baryons are found in the gaseous interstellar medium of these systems. It is therefore unlikely that our metallicities are representative of the true global metallicity \citep{2014ApJ...795..109J}. Furthermore, galaxies that have formed a substantial fraction (i.e., $>$10$\%$) of their stars in a recent star formation episode often have $M/L$ ratios that deviate from typical $M/L$ ratios. Although we have taken caution to use SDSS bands least likely to be contaminated by the ongoing or recent star formation event, even NIR stellar $M/L$ ratios can vary, depending on factors such as star formation rate and metallicity \citep{2001ApJ...550..212B}.

Overall, however, our sample of BCDs, particularly those with direct abundance measurements, follow the \citet{2012ApJ...754...98B} $M-Z$ relation slightly more closely than they do the $L-Z$ relation, with a mean residual from Equation \ref{equation:MZ} of 0.264. This supports existing studies that the $M-Z$ relation is the more fundamental of the two relations. 

\subsection{The Search for BCDs in other Photometric Surveys}
\label{BCDs_othersurveys}
With the advent of numerous photometric surveys, our presented method of identifying candidate low metallicity galaxies via photometry alone can be adapted to query the data products of forthcoming astronomical surveys to further increase the number of local galaxies with metallicities less than 12\,+\,log(O/H)\,$\leq$\,7.65. Multiple ongoing surveys such as PanSTARRS, the Dark Energy Survey (DES), and the Dark Energy Camera Legacy Survey (DECaLs) can each supplement the photometric search for low metallicity systems and offer the following advantages: both PanSTARRS and DES will survey larger areas of the sky than covered by SDSS, and in particular, the DES footprint will scan the southern hemisphere, providing photometric information of sky regions not covered by current surveys. DECaLS will reach fainter magnitudes and potentially uncover low metallicity systems in our local Universe that are currently below the detection limit of SDSS. Additionally, these surveys can extend the search for low metallicity systems to somewhat higher redshifts. As shown in Figure \ref{fig:redshift}, our BCD sample has a mean redshift of $z$\,=\,0.016 and reaches a maximum redshift of $z$\,=\,0.052. Oncoming surveys that reach higher redshifts can therefore cover a much greater volume (i.e., a survey that can reach twice as far as current limits would probe eight times the current volume).

However, searching for low metallicity galaxies in either PanSTARRS, DES, or DECaLS is complicated by the lack of $u$-band photometry, particularly because the most metal-poor systems currently known in the local Universe appear to cluster around a tight $u-g$ color space, as shown in Figure \ref{fig:colorcuts}. Our current SDSS query parameters will require modification to efficiently pick out the same objects in their various color-color spaces -- $grizy$ in PanSTARRS, $grizY$ in DES, and $grz$ in DECaLS. We note that the Canada France Imaging Survey (CFIS; \citealt{2017ApJ...848..128I}) offers $u$-band photometry and an overlap in footprint with the DES, allowing the two to be used in conjunction.  Finally, by extending the search for low metallicity dwarf galaxies to a larger volume, the change in photometric colors as we move into higher redshifts must also be taken into account.

\section{Conclusion}
\label{C}
We present spectroscopic observations of 94 newly identified BCDs using the Kast spectrograph on the Shane 3-m telescope at Lick Observatory and LRIS at the W.M. Keck Observatory. The BCDs were first identified as candidate low-metallicity systems via their photometric colors in Data Release 12 of the Sloan Digital Sky Survey. From this query, we selected a subset of objects best fit for observing based on their morphologies.

From our observations, we estimate the gas-phase oxygen abundances of our observed systems using the $R$ and $S$ calibrations for objects observed using the Kast spectrograph and make direct oxygen abundance measurements for systems observed using LRIS, where the temperature-sensitive [\OIII]$\lambda$4363\,\AA\ line is detected. %Almost half of our observations are in the low metallicity regime, with metallicities $\lesssim$\,0.1\,Z$_{\odot}$ or 12\,+\,log(O/H)\,$\leq$\,7.65.

These observations are part of a recent survey led by the authors to identify low metallicity systems based on photometry alone. To date, this program has yielded highly successful results in discovering new metal-poor systems. Specifically, our initial observations of candidate BCDs yielded 67\% of systems to be emission-line galaxies. Of the confirmed emission line sources, 45\% are in the low metallicity regime, with metallicities $\lesssim$\,0.1\,Z$_{\odot}$ or 12\,+\,log(O/H)\,$\leq$\,7.65, and 6\% have been confirmed or are projected to be in the lowest metallicity regime, 12\,+\,log(O/H)\,$\leq$\,7.20. %(1) emission line sources: Kast success, including those later followed-up on using LRIS+LRIS success = (85/135)+(16/16) = 67%; (2) 12+log(O/H)<=7.65 = 45/(85+16); (3) 12+log(O/H)<=7.2 = 6/(85+16)
This technique is a promising means of bolstering the current meager number of systems that push on the low-luminosity and lowest metallicity regime. Using photometry to identify candidate low-metallicity systems can provide a more efficient yield in finding extremely metal-poor systems in comparison to existing programs, which have mostly relied on existing spectroscopic information, from which metal-poor systems are then identified.

With new data from ongoing and upcoming all-sky photometric surveys that add new sky coverage and reach deeper magnitudes, our method promises to greatly increase the number of known low metallicity systems, particularly pushing on the lowest metallicity regime, where only a handful of systems are currently known with 12\,+\,log(O/H)\,$\leq$\,7.20, and reaching a larger volume of the Universe.

\acknowledgments
We are grateful to our anonymous referee for their thorough comments which have resulted in an improved manuscript. The authors thank Alis Deason and Connie Rockosi for useful discussions, particularly in the development of this survey. We thank Evan Skillman and Erik Aver for their expertise and helpful input throughout the analysis of our sample, particularly in correcting for reddening and underlying stellar absorption. The data presented herein were obtained at the W.M. Keck Observatory, which is operated as a scientific partnership among the California Institute of Technology, the University of California and the National Aeronautics and Space Administration. The Observatory was made possible by the generous financial support of the W.M. Keck Foundation. The authors wish to recognize and acknowledge the very significant cultural role and reverence that the summit of Mauna Kea has always had within the indigenous Hawaiian community.  We are most fortunate to have the opportunity to conduct observations from this mountain. Research at Lick Observatory is partially supported by a generous gift from Google. The red-side CCD upgrade of the Lick Observatory Kast Spectrometer was made possible by generous gifts from the Bill and Marina Kast and the Heising-Simon's Foundation. We also gratefully acknowledge the support of the staff at Lick and Keck Observatories for their assistance during our observing runs. During this work, R.~J.~C. was supported by a Royal Society University Research Fellowship, and by NASA through Hubble Fellowship grant HST-HF-51338.001-A, awarded by the Space Telescope Science Institute, which is operated by the Association of Universities for Research in Astronomy, Inc., for NASA, under contract NAS5-26555. R.~J.~C. acknowledges support from STFC (ST/L00075X/1). J.~X.~P. acknowledges support from the National Science Foundation grant AST-1412981.

\facilities{Shane (Kast Double spectrograph), Keck:I (LRIS)}
\software{\texttt{astropy} \citep{2013A&A...558A..33A},\, \texttt{matplotlib} \citep{Hunter:2007},\, \texttt{NumPy}\,\citep{2011arXiv1102.1523V},\,\texttt{SciPy}\, \citep{SciPy}}

%\bibliographystyle{yahapj}
%\bibliography{references}

\appendix
\section{SDSS CasJobs Query}
\label{CasJobsquery}
\begin{center}
\texttt{
\textsc{SELECT} P.ObjID, P.ra, P.dec, P.u, P.g, P.r, P.i, P.z \textsc{into} mydb.MyTable \textsc{from} Galaxy P \\
\textsc{WHERE} \\
(\,P.u\,-\,P.g\,$>$\,0.2\,) \\
\textsc{and} (\,P.u\,-\,P.g\,$<$\,0.60) \\
\textsc{and} (\,P.g\,-\,P.r\,$>$\,-0.2\,) \\
\textsc{and} (\,P.g\,-\,P.r\,$<$\,\,0.2\,) \\
\textsc{and} (\,P.r\,-\,P.i\,$<$\,-0.1\,) \\
\textsc{and} (\,P.r\,-\,P.i\,$>$\,-0.7\,) \\
\textsc{and} (\,P.i\,-\,P.z\,$<$\,0.1\,) \\
\textsc{and} (\,P.i\,-\,P.z\,$>$\,-0.4\,-\,2$*$P.err\textunderscore z\,) \\
\textsc{and} (\,P.r\,$<$\,21.5) \\
\textsc{and} ((\,P.b\,$<$\,-25.0)\,or\,(\,P.b\,$>$\,25.0)) \\
\textsc{and} (P.fiberMag\textunderscore g\,$<$\,P.fiberMag\textunderscore z)}
\end{center}

\end{document}